\documentclass[
12pt]{article}
\usepackage{color}
\usepackage{graphicx}
 \usepackage{amsmath,amssymb,array,calc}

\definecolor{darkgreen}{rgb}{0,0.55,0}

\newcommand{\bea}{\begin{eqnarray}}
\newcommand{\eea}{\end{eqnarray}}
\newcommand{\be}{\begin{equation}}
\newcommand{\ee}{\end{equation}}
\newcommand{\nn}{\nonumber}

\def\revise#1       {\raisebox{-0em}{\rule{3pt}{1em}}%
                     \marginpar{\raisebox{.5em}{\vrule width3pt\
                     \vrule width0pt height 0pt depth0.5em
                     \hbox to 0cm{\hspace{0cm}{%
                     \parbox[t]{4em}{\raggedright\footnotesize{#1}}}\hss}}}}

\def\Re           {{\rm Re\hskip0.1em}}
\def\Im           {{\rm Im\hskip0.1em}}

\def\sqr#1#2{{\vcenter{\vbox{\hrule height.#2pt
 \hbox{\vrule width.#2pt height#1pt \kern#1pt
 \vrule width.#2pt}\hrule height.#2pt}}}}

\catcode`\@=12

\begin{document}

\makeatletter \@addtoreset{equation}{section} \makeatother
\renewcommand{\theequation}{\thesection.\arabic{equation}}

\renewcommand\baselinestretch{1.25}
\setlength{\paperheight}{10in}
\setlength{\paperwidth}{9.5in}

\begin{titlepage}


%

\vskip 1cm



\vskip 3cm

 \begin{center}
{\bf \large Real-time finite-temperature correlators from AdS/CFT}
\end{center}
\vskip 1cm

\centerline{{\large  Edwin Barnes, Diana Vaman, Chaolun Wu and Peter Arnold}\footnote{E-mail addresses: eb4df,dv3h,cw2an,parnold@virginia.edu}
}

\vskip .5cm
\centerline{\it Department of Physics, The University of Virginia}
\centerline{\it McCormick Rd, Charlottesville, VA 22904}

\vspace{1cm}

\begin{abstract}

In this paper we use AdS/CFT ideas in conjunction with insights from finite temperature real-time field theory formalism to compute 3-point correlators of ${\cal N}{=}4$ super Yang-Mills operators, in real time and at finite temperature. To this end, we propose that the gravity field action is integrated only over the right and left quadrants of the Penrose diagram of the Anti de Sitter-Schwarzschild background, with a relative sign between the two terms. For concreteness we consider the case of a scalar field in the black hole background. Using the scalar field Schwinger-Keldysh bulk-to-boundary propagators, we give the general expression of a 3-point real-time Green's correlator.  We then note that this particular prescription amounts to adapting the finite-temperature analog of Veltman's circling rules to tree-level Witten diagrams, and comment on the retarded and Feynman scalar bulk-to-boundary propagators.  We subject our prescription to several checks: KMS  identities,  the largest time equation and the zero-temperature limit. When specializing to a particular retarded (causal) 3-point function, we find a very simple answer: the momentum-space correlator is given by three causal (two advanced and one retarded) bulk-to-boundary propagators, meeting at a vertex point which is integrated from spatial infinity to the horizon only. This result is expected based on analyticity, since the retarded n-point functions are obtained by analytic continuation from the imaginary time Green's function, and based on causality considerations.

\end{abstract}

\end{titlepage}



\section{Introduction and Summary}

In recent years, there has been a great deal of interest in applying AdS/CFT methods to the study of strongly-coupled quark-gluon plasmas created at the Relativistic Heavy Ion Collider (RHIC) at Brookhaven National Laboratory. This interest stems from the fact that some features of strongly-coupled quark-gluon plasmas are captured by an analytically tractable example of a gauge theory plasma at strong coupling: ${\cal N}{=}4$ $SU(N_c)$ super Yang-Mills theory at finite temperature.  In the deconfined phase, this theory is expected to qualitatively resemble high-temperature QCD (see e.g. \cite{Herzog:2006gh} for a discussion of this). In the limit of large $N_c$ and large 't Hooft coupling, ${\cal N}{=}4$ super Yang-Mills is holographically dual to weakly-coupled string theory on a black hole spacetime which asymptotes to the product of 5-dimensional Anti-de Sitter space (AdS) with a 5-dimensional sphere \cite{Maldacena:1997re, Witten:1998qj, Gubser:1998bc, Witten:1998zw}. Using the AdS/CFT correspondence, many dynamical properties of the ${\cal N}{=}4$ strongly coupled plasma have been investigated: e.g. transport coefficients such as viscosity were related to a computation of real-time 2-point correlators in the black hole background (see \cite{ss} and references within). The prescription for computing 2-point real-time correlators goes back to Son and Starinets \cite{Son:2002sd} and to Son and Herzog \cite{Herzog:2002pc} (for more recent work see \cite{Skenderis:2008dh}). On the other hand, very little is known about computing 3-point and higher correlators of operators in real time and at finite temperature. In fact, this paper gives the very first concrete expression for real-time finite-temperature 3-point correlators.

According to AdS/CFT, one can compute a 3-point correlator in super Yang-Mills by computing the analog of a Feynman diagram (Witten diagram) in the Anti-de Sitter-Schwarzschild (AdS-S) background. This diagram includes three bulk-to-boundary propagators for the gravity field which couples to the super Yang-Mills operator under consideration, and these propagators meet at a vertex point in the bulk. The location of the vertex must be integrated over. For real-time (Minkowski signature) calculations, a possible source of confusion concerns the range of integration. Should one integrate only up to the horizon in Schwarzschild coordinates? Or should one perhaps think about the Penrose diagram of the AdS-S space and integrate over the right and left quadrants only? One could instead integrate over the whole diagram (including the past and future quadrants), but then one would need to specify how to carry out the integration at the singularities. Or maybe one has to use an entirely different gravitational background by gluing different geometries in such a way that one constructs a gravitational analogue of the Schwinger-Keldysh contour. Various authors have proposed different ways of addressing this issue  \cite{Herzog:2002pc}, \cite{Skenderis:2008dh, Skenderis:2008dg}. We will resolve this integration ambiguity by extending the work of Son and Herzog and using many insights from the formalism of real-time finite-temperature field theory, offering an alternative to other approaches in the literature which is quite simple to implement, and which yields consistent results.

For simplicity, we consider operators which couple to (massless) scalar fields in the AdS-S background, but we expect that our results generalize easily to other spins. We first construct the Schwinger-Keldysh scalar bulk-to-boundary propagator. This is a $2\times 2$ matrix whose indices correspond to the location of the boundary point (either the right (R) or left (L) boundary) and to the location of the bulk vertex point (either in the R or L quadrants). The minimally coupled scalar action is defined to be integrated over only the R and L quadrants, with a relative sign between the two bulk integrals. This is along the lines of Frolov and Martinez \cite{Frolov}, who in turn were inspired by Israel's thermofield formalism \cite{Israel:1976ur}. It turns out that the relative sign between the two contributions precisely matches the relative sign between the contributions of the physical and doubler fields in real-time finite-temperature field theory. We note, following Kobes and Semenoff \cite{Kobes:1985kc}, that the Schwinger-Keldysh propagator can be re-interpreted in terms of circling rules (that is, the entries of the Schwinger-Keldysh propagator can be related to the Feynman and Wightman propagators, which are represented diagrammatically in terms of Veltman's circling rules \cite{Veltman:1963th, 'tHooft:1973pz}). Thus the R-L prescription can be traded for circling rules in the AdS-S background, and with the integration over the bulk vertex only up to the horizon. The various real-time 3-point functions are then computed diagrammatically by placing circles around the boundary and bulk vertex points, and summing the appropriate number of diagrams. For example, a time-ordered product of three operators in real time and finite temperature is computed by adding two gravity diagrams: one with three boundary vertices of type 1 (uncircled) and three bulk-to-boundary propagators meeting at a bulk vertex which is either of type 1 (uncircled) or 2 (circled). As we have already mentioned, the bulk vertex is integrated only up to the horizon.

We subject our prescription to several checks: Kubo-Martin-Schwinger identities, the largest time equation, and the zero-temperature limit. Moreover, by focusing on a special 3-point real-time correlator which is {\it retarded}, we can make contact with the 3-point correlator obtained in imaginary time (Euclidean signature). The latter can be computed straightforwardly using Witten-type diagrams in Euclidean AdS-S, without encountering any of the subtleties we discussed before: Euclidean AdS-S has no singularities (it caps at the horizon) and has only a single boundary, just like Euclidean AdS. The retarded 3-point function is related to the imaginary-time 3-point Green's function by analytic continuation in frequency. Our more general prescription for computing generic real-time finite-temperature 3-point correlators yields a retarded 3-point function which has this property. In the past, analyticity arguments were used in the context of AdS/CFT by Gubser et al. \cite{Gubser:2008sz} and by Iqbal and Liu \cite{Iqbal:2008by, Iqbal:2009fd} to obtain the real-time 2-point correlators. Lastly, using the causal nature of the retarded 3-point, we can understand why the integration region over the bulk vertex is only up to the horizon, that is, why one needs to integrate only over the R region of the Penrose diagram (similar causality arguments were used by Caron-Huot and Saremi \cite{CaronHuot:2009iq} when computing one-loop gravity corrections to the retarded 2-point correlator).

The paper is organized as follows: In Section 2, we first derive the real-time 3-point correlators at zero temperature. This is done so that we have a reference point when asking the question: is the finite-temperature prescription compatible with the zero-temperature result? In our approach we start from the position-space field theory correlators and Fourier-transform. Amusingly, in this procedure, the AdS radial coordinate arises as a Schwinger parameter.
Next we find that the real-time correlators are obtained by integrating over a region of the AdS space, the Poincar\'{e} patch. This is in contrast to the more familiar AdS/CFT story in Euclidean signature, where the AdS tree-level diagrams are obtained by integrating over the whole AdS space.
In particular, the final answer for the momentum-space retarded 3-point correlator is expressed as the product of three causal (two advanced and one retarded) bulk-to-boundary AdS propagators, integrated over the position of the bulk vertex only over the Poincar\'{e} patch.

Then, in Section 3 we review some of the basic elements of real-time finite-temperature field theory formalism. In particular we take note of an observation made by Kobes \cite{kobes} that in real-time formalism, the retarded n-point function, which he proceeds to define, is the real-time correlator which is obtained by analytic continuation in frequency space from the imaginary-time finite-temperature correlator.

In Section 4, we spend some time reviewing the construction of the bulk-to-boundary propagators. In particular, the retarded bulk-to-boundary scalar propagator is the one which behaves like an incoming wave at the horizon \cite{CaronHuot:2009iq}. We give a self-contained exposition of its expression in terms of Heun's functions, and explore its analytic properties, connection with the Euclidean signature bulk-to-boundary propagator, causal properties and zero-temperature limit. Then we notice that the Feynman propagator in curved space is associated with a choice of vacuum.  We construct the Feynman Green's function from the retarded Green's function by borrowing the definition of a thermal Feynman Green's function from finite-temperature field theory:
\be
{\cal G}_F(E,\vec P,u)=\Re {\cal G}_R(E,\vec P,u)+i\Im {\cal G}_R (E,\vec P,u)
{\rm coth}(\beta E/2)
\label{fdt},
\ee
where $u$ is related to the radial/holographic direction, and $\beta$ is the inverse temperature. We then note that the Schwinger-Keldysh bulk-to-boundary propagator can be re-expressed in terms of circling rules propagators \cite{Veltman:1963th, 'tHooft:1973pz}  in a way entirely analogous to real-time finite-temperature field theory \cite{Kobes:1985kc}.

We begin Section 5 by reviewing the real-time finite-temperature 2-point correlators from AdS/CFT. Our definition for the scalar field action is in accord with the R-L prescription. This choice is such that by evaluating the on-shell AdS-S scalar kinetic action and picking up the R and L boundary terms, one gets the 2-point Schwinger-Keldysh field theory correlator. Next, we use the same R-L prescription to evaluate the 3-point real-time finite-temperature correlators. We have already advertised the result: it is consistent with the known identities obeyed by the real-time finite-temperature correlators, and with the zero-temperature limit. Then, by focusing on the retarded finite-temperature 3-point correlator, we find that although numerous terms contribute, the answer is quite simple: it can be viewed as arising from a single Witten-type diagram containing three causal propagators (two advanced and one retarded) joined at a bulk vertex which is integrated only up to the horizon. This is what one expects to get based on considerations of the zero-temperature limit and analytic continuation from Euclidean signature. Moreover, for a causal 3-point function, one expects that one has to integrate only over a causal bulk region determined by the boundary points. Since our result gives the momentum-space 3-point retarded correlator, the bulk integration can only be over the R quadrant, which is a maximal causal diamond. We view all of these results as consistency checks of our prescription for the computation of the real-time finite-temperature 3-point functions, which can also be summed up by the use of circling rules, and a bulk integration region from spatial infinity up to the black hole horizon.

The technical details are relegated to Appendices. In Appendix A, for completeness, we give the momentum-space zero-temperature 2-point CFT correlators. In Appendix B, we obtain the real-time momentum-space zero-temperature 3-point functions in terms of Witten-type diagrams by employing a reverse-engineering procedure, namely starting from the CFT correlators and re-packaging them as Witten diagrams. In Appendix C, we discuss the zero-temperature retarded bulk-to-boundary scalar propagators. In Appendix D, we give the closed-form expression of the zero-temperature retarded 3-point function for CFT operators with conformal dimension $\Delta=2$. Lastly, in Appendix E we give some background material on the Heun function and its various local representations.

\section{From field theory $T{=}0$ real-time correlators to AdS Witten diagrams}

\subsection{Real-Time CFT correlators}
To set our notation in a simplified setting, let us begin with a massless scalar field theory. In Minkowski signature (that is, real time), one encounters different Green's functions. Besides the Feynman propagator,
\bea
D_F (x) &\equiv& \langle 0 |{\cal T} \phi(x) \phi(0) |0\rangle \nonumber\\
&\equiv& \theta(t) D^+(x) + \theta(-t) D^-(x)\nonumber\\
& =&
\frac{1}{(2\pi)^2}\frac{1}{(-t^2 + \vec{x}^2 + i\epsilon)},
\eea
where $D^\pm$ are Wightman's functions
\bea
D^-(x)&=&\langle 0 | \phi(0) \phi(x) | 0\rangle\nonumber\\
&=& \frac{1}{(2\pi)^2}\frac{1}{[-(t+i\epsilon)^2+\vec x^2]},\\
D^+(x)=&=&\langle 0 | \phi(x) \phi(0) | 0\rangle\nonumber\\
&=&\frac{1}{(2\pi)^2}\frac{1}{[-(t-i\epsilon)^2+\vec x^2]},
\eea
one defines the causal retarded and advanced propagators,
\bea
D_R(x) &\equiv & \theta(t)\langle [\phi(x),\phi(0)] \rangle\nonumber\\
&=& D_F(x) - D^-(x) 
\nonumber\\
&=&-\frac{i}{2\pi} \theta(t) \delta(-t^2 + \vec x^2)
\\
D_A(x)  &\equiv& -\theta(-t)\langle [\phi(x),\phi(0)] \rangle\nonumber\\
&=&D_F(x)-D^+(x)\nonumber\\&=& -\frac{i}{2\pi} \theta(-t) \delta(-t^2 + \vec x^2) .
\eea
The retarded/advanced propagators  (and the higher n-point functions) can  be obtained following Veltman's circling rules \cite{Veltman:1963th, 'tHooft:1973pz}. According to these rules, each vertex can be circled or uncircled. The circling of a vertex brings a minus sign. The propagator between uncircled vertices is the usual Feynman propagator, $D_F(x)$. The propagator between two circled vertices is the complex conjugate of the Feynman propagator, $D_F^*(x)$. The propagator between a circled vertex and an uncircled one is $D^-(x)$, and the propagator between an uncircled vertex and a circled one is $D^+(x)$ (see Fig.1).
\begin{figure}
\begin{center}
\includegraphics[width=1.3in]{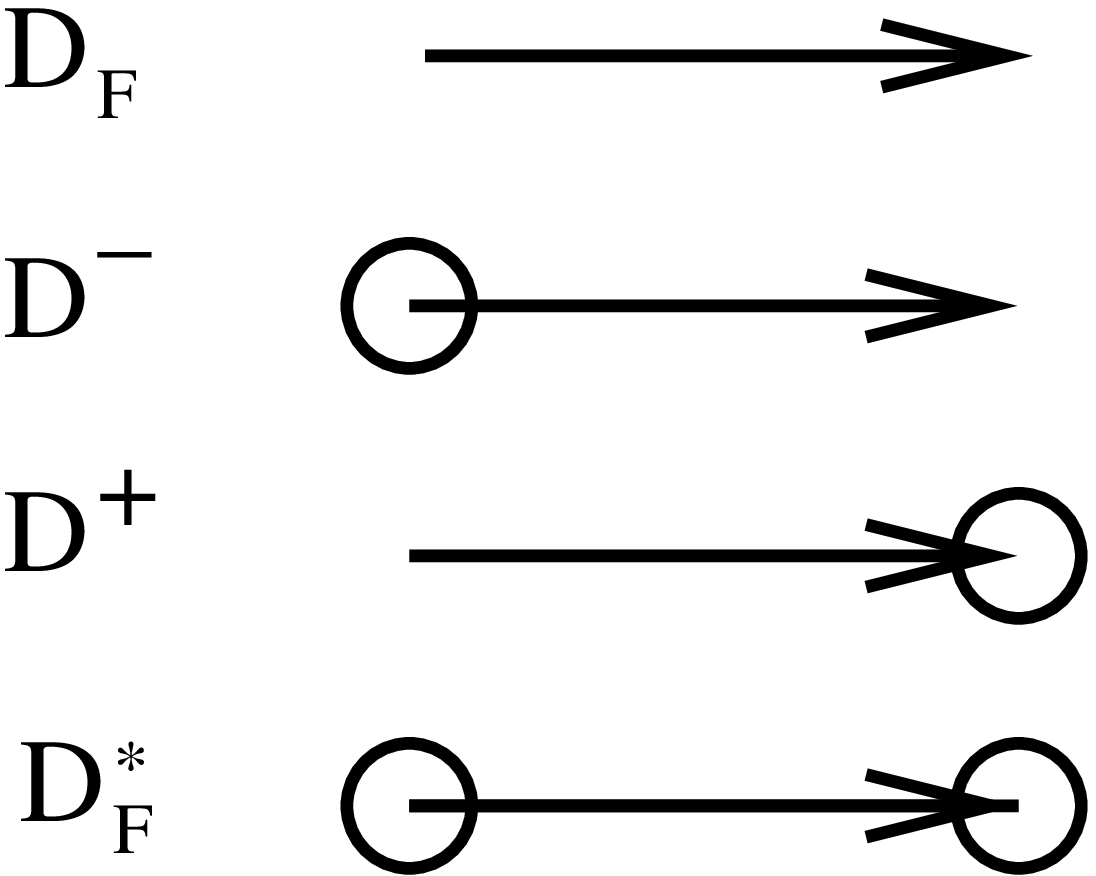} \\
Figure 1: Circling rules propagators
\end{center}
\end{figure}
As a consequence of these circling rules, Veltman was able to formulate the largest time equation \cite{Veltman:1963th}. This is an algebraic identity stating that the sum of all diagrams obtained from a single Feynman diagram by placing circles around all vertices in all possible combinations, for a total of $2^n$ diagrams if there are $n$ vertices, is zero. For example, one can easily verify that $D_F - D^- - D^++ D_F^*=0$.

The causal retarded n-point function where the largest time is associated with a certain vertex is computed by adding all diagrams, where each vertex can be circled or uncircled, with the exception of the vertex associated with the largest time which remains uncircled. For the 2-point, this indeed reduces to $D_R(x) = D_F(x) - D^-(x)$.

Our notation is such that the propagator will be consistently denoted by fonts of the letter $D$, while the n-point Green's functions will be denoted by fonts of the letter $G$. For example, the 2- and n-point Feynman Green's functions are given by
\be
iG_F(x,x')= \langle {\cal T} \phi(x)\phi(x')\rangle, \qquad
i^{n-1}G_F(x_1,x_2\dots x_n)= \langle {\cal T} \phi(x_1)\phi(x_2)\dots \phi(x_n)\rangle.
\ee

In a conformal field theory, the 2-point function is fixed by symmetry. Considering for simplicity scalar operators of the same conformal dimension $\Delta$, the various real-time 2-point functions are\footnote{
The ${\cal N}{=}4$ super Yang-Mills theory 2-point functions have an additional overall constant $8\frac{\Delta-2}{\Delta}\frac{\Gamma(\Delta+1)}{\Gamma(\Delta-2)}$ which was stripped off from the subsequent formulae, together with a factor which is dependent on the gauge group and representation.
}:
\bea
G_F(x)&=&-i \langle 0 | {\cal T} O(x) O(0) | 0 \rangle=  -\frac{i}{(2\pi)^2}
\bigg(\frac{1}{-t^2 + \vec x^2 + i\epsilon}\bigg)^\Delta,\\
 G^+(x) &=& -i \langle 0 | O(x) O(0) | 0\rangle= \theta(t) G_F(x) - \theta(-t) G_F^*(x) = -\frac{i}{(2\pi)^2} \bigg(\frac{1}{-(t-i\epsilon)^2+ \vec x^2}\bigg)^\Delta,
\nonumber\\
\\
G^-(x) &=& -i \langle 0 | O(0) O(x) | 0 \rangle= \theta(-t) G_F(x) - \theta(t) G_F^*(x)=-\frac{i}{(2\pi)^2} \bigg(\frac{1}{-(t+i\epsilon)^2+ \vec x^2}\bigg)^\Delta,
\nonumber\\
\\
G_R(x) &=& G_F(x) - G^-(x) = \theta(t) \bigg(G^+(x) - G^-(x)\bigg),\\
G_A(x) &=& G_F(x) - G^+(x) =\theta(-t) \bigg(G^-(x) - G^+(x)\bigg).
\eea
The CFT Green's functions have the same $i\epsilon$ prescription as the scalar Green's functions. For completeness, we give the momentum-space 2-point functions in Appendix A. An important observation which carries through, as will see when discussing the retarded momentum-space propagator of scalar fields in a black hole background, is that the analytic continuation of the Euclidean signature 2-point function, with $E\to -i(E\pm i\epsilon) $ yields the retarded/advanced 2-point correlators \cite{Gubser:2008sz}  \cite{Iqbal:2008by,Iqbal:2009fd}.

Moving on to 3-point functions, these are also fixed by conformal symmetry (up to an overall constant, which was set to 1). For example
\bea
G_F(x_1,x_2,x_3)&=&(-i)^2\langle  0 | {\cal T} O(x_1) O(x_2) O(x_3) | 0 \rangle\nonumber\\
&=&(-i)^2\bigg(\frac{1}{(-t_{12}^2 + \vec x_{12}^2 + i\epsilon)} \,\frac{1}{(-t_{23}^2 + \vec x_{23}^2 + i\epsilon)}\,\frac{1}{(-t_{31}^2 + \vec x_{31}^2 + i\epsilon)} \bigg)^{\Delta/2} .
\eea
The non-time-ordered product of three scalar operators is
\bea
G_{123}(x_1,x_2,x_3)&=&(-i)^2\langle 0| O(x_1) O(x_2) O(x_3) | 0 \rangle
\nonumber\\&=&(-i)^2\bigg(\frac{1}{(-t_{12}^2 + \vec x_{12}^2 + i\epsilon t_{12})} \,\frac{1}{(-t_{23}^2 + \vec x_{23}^2 + i\epsilon t_{23})}\,\frac{1}{(-t_{31}^2 + \vec x_{31}^2 - i\epsilon t_{31})} \bigg)^{\Delta/2}, \nonumber\\
\eea
where the $i\epsilon$ prescription follows the same rules as for the 2-point functions, namely the time coordinate of an operator insertion is greater by $i\epsilon$ than the time coordinate of any other operator insertion to the right of it \cite{wightman} (for a recent AdS/CFT paper making use of this $i\epsilon$ prescription see \cite{Hofman:2008ar}).

Knowing these correlators means that we can obtain the retarded 3-point function for which $x_3$ has the largest time by using the analytic continuations which follow from the circling rules (alternatively, see \cite{ls}):
\bea
\!\!\!\!\!\!\!\!\!\!\!\!&&\!\!\!\!\!\!\!\!\!\!\!\!G_R(x_1,x_2;x_3)=
\theta(t_{31})\theta(t_{12})\bigg(G_{312}-G_{132}+G_{213}-G_{231}\bigg)
\nonumber\\
&&\;\;\;\;\;\;\;\;\;\;\;\;\;\;\;+
\theta(t_{32})\theta(t_{21})\bigg(G_{321}-G_{231}+G_{123}-G_{132}\bigg)
\nonumber\\
&=&(-i)^2\theta(t_{31})\theta(t_{12})\bigg[\bigg(\frac{1}{(x_{12}^2-i\epsilon t_{12})^\delta}\frac{1}{(x_{23}^2+i\epsilon t_{23})^\delta}-c.c.\bigg)\bigg(
\frac{1}{(x_{31}^2-i\epsilon t_{31})^\delta}-\frac{1}{(x_{31}^2+i\epsilon t_{31})^\delta}
\bigg)\bigg]\nonumber\\
&+&(-i)^2\theta(t_{32})\theta(t_{21})
\bigg[\bigg(\frac{1}{(x_{12}^2+i\epsilon t_{12})^\delta}\frac{1}{(x_{31}^2-i\epsilon t_{31})^\delta}-c.c.\bigg)\bigg(
\frac{1}{(x_{23}^2+i\epsilon t_{23})^\delta}-\frac{1}{(x_{23}^2-i\epsilon t_{23})^\delta}
\bigg)
\bigg],\nonumber\\\label{retarded3point}
\eea
where we defined for brevity
\be
\delta\equiv\frac{\Delta}2,
\ee
and $G_{ABC}=(-i)^2\langle O(x_A) O(x_B) O(x_C)\rangle $ are the non-time-ordered correlators.

\subsection{Lessons for real-time AdS/CFT from CFT correlators}
Before we discuss the real-time correlators, we make a detour in Euclidean space, and show how one could discover Witten's AdS/CFT diagrams starting from the 3-point momentum-space CFT correlator. We begin by Fourier-transforming the position-space 3-point correlator. To perform the integrals, we introduce three Schwinger parameters. The integrals over the Euclidean-signature coordinates $x^\mu$ are of Guassian type, and easily performed. To deal with the intermediate result, we need to introduce one more Schwinger parameter, $z$. As shown in detail in Appendix B, the final answer is:
\bea
&&\!\!\!\!\!\!  i^2G_E(k_1,k_2,k_3)=\nonumber\\
&=&2^{10-3\Delta{-3\epsilon}}(2\pi^2)^{4-2\epsilon} \frac{\delta^{4-2\epsilon}(k_1+k_2+k_3)}{ \Gamma(\frac{\Delta}{2})^3\Gamma(\frac{3\Delta}2-\frac{4-2\epsilon}{2})}\int_0^\infty \frac{dz}{z^{5-2\epsilon}}
\prod_{i=1}^3 (\sqrt{k_i^{2}})^{\Delta-\frac{4-2\epsilon}{2}}z^{\frac{4-2\epsilon}{2}}
K_{\Delta-\frac{4-2\epsilon}{2}}(\sqrt{k_i^2}z)
\nonumber\\\label{3pointE}
\eea
The expression in (\ref{3pointE}) is UV-divergent, and to regularize it we employed here dimensional regularization.
It is amusing to notice that the momentum-space 3-point CFT correlator has been reassembled as an integral over the AdS radial coordinate $z$ of the product of three bulk-to-boundary AdS scalar propagators. This is precisely a 3-point function Witten diagram \cite{Witten:1998qj}.
With a bit of hindsight, from the measure factor and from the form of the propagators in (\ref{3pointE}), the AdS metric can be reconstructed to yield
\be
ds_5^2 = \frac{dz^2+ dx^\mu dx_\mu}{z^2},\label{poincare}
\ee
In Euclidean signature the Poincar\'{e} coordinates of (\ref{poincare}) cover the whole AdS space.

We should also point out that as a result of our use of dimensional regularization in computing (\ref{3pointE}), the AdS space has dimension ${d+1}$, with $d=4-2\epsilon$. The scalar field with mass $m$ and d-dimensional momentum $k^\mu$ in $(d+1)$-dimensional Euclidean AdS space has a bulk-to-boundary propagator
\be
(\sqrt{k^{2}})^{\Delta-\frac{d}{2}}z^{\frac{d}{2}}
K_{\Delta-\frac{d}{2}}(\sqrt{k^2}z), \label{euclt0prop}
\ee
where $\Delta$  is the conformal dimension of the CFT operator which couples to mass $m$ scalar field:
\be
\Delta=\frac{1}{2}(d+\sqrt{d^2+4m^2}).
\ee

A different regularization which is more commonly used in AdS/CFT calculations imposes boundary conditions  at $z=z_B\ll1$ \cite{Freedman:1998tz}. Then, the bulk-to-boundary propagators which have a $\delta$-function support at $z_B$ in position space, have the following expression in momentum space: 
\be
\frac{ z^{\frac d2} K_{\Delta - \frac d2}(\sqrt{k_i^2}z)}{z^{\frac d2}_B K_{\Delta- \frac d2}(\sqrt{k_i^2}z_B)}.\label{normalization}
\ee

Next, we would like to compute the Fourier transform of (\ref{retarded3point}) to get the momentum  space retarded 3-point correlator.
The reason to perform this calculation is to find out how the Witten diagrams look like in real-time. In particular we want to find out which is the region of the Minkowski-signature AdS space where the tree-level gravity diagrams need to be integrated over.
This becomes relevant later, in Section 5, when we will inquire whether the real-time finite-temperature 3-point correlators have the correct zero-temperature limit.

As we will see, the answer is quite natural: since we want to identify the 4-dimensional coordinates
$x^\mu$ in the field theory and in the holographic dual, we will be using the Poincar\'{e} parametrization of the AdS metric (\ref{poincare}) with Minkowski signature. This also comes out naturally from the reverse engineering perspective that we are following: as in (\ref{3pointE}), this identification is built-in. It follows that we will not be integrating over the whole AdS space, since (\ref{poincare}) now covers just half of AdS. As for the question of whether in real-time AdS/CFT we will be integrating over the whole range of the radial coordinate $z$, we can anticipate that the answer will be affirmative, if  the radial coordinate enters, as it did before, as a Schwinger parameter.
Lastly, we might expect that the retarded 3-point is obtained by performing an analytic continuation $\omega\to-i(\omega+i\epsilon)$ of the Euclidean result, and that perhaps this analytic continuation is allowed under the integral (\ref{3pointE}). The answer to this second question is again affirmative.

The most straightforward proof of our previous statements requires that we choose a certain momentum kinematics for the momentum-space correlator: if the time associated with the spacetime point $x_3$ is the largest of $x_1,x_2,x_3$, we will take the momenta $p_{1,2,3}^\mu$ to be such that $E_{1,2}<0, E_3>0$.
 With these kinematics, we can show that our retarded 3-point correlator in momentum space is given by the Fourier transform of $G_F(x_1,x_2,x_3)$. To see this we only need to repackage the retarded 3-point in position space as
\bea
i^2G_R(x_1,x_2;x_3)&=&\bigg(\frac{1}{-t_{12}^2 + \vec x_{12}^2 + i\epsilon} \,\frac{1}{-t_{23}^2 + \vec x_{23}^2 + i\epsilon}\,\frac{1}{-t_{31}^2 + \vec x_{31}^2 + i\epsilon} \bigg)^{\Delta/2}\nonumber\\ &-&
\bigg(\frac{1}{-(t_{12}-i\epsilon)^2 + \vec x_{12}^2} \,\frac{1}{-t_{23}^2 + \vec x_{23}^2 + i\epsilon}\,\frac{1}{-(t_{31}+i\epsilon)^2 + \vec x_{31}^2 } \bigg)^{\Delta/2}\nonumber\\ &-&
\bigg(\frac{1}{-(t_{12}+i\epsilon)^2 + \vec x_{12}^2} \,\frac{1}{-(t_{23}-i\epsilon)^2 + \vec x_{23}^2}\,\frac{1}{-t_{31}^2 + \vec x_{31}^2 + i\epsilon} \bigg)^{\Delta/2}\nonumber\\ &+&
\bigg(\frac{1}{-t_{12}^2 + \vec x_{12}^2 - i\epsilon} \,\frac{1}{-(t_{23}-i\epsilon)^2 + \vec x_{23}^2 }\,\frac{1}{-(t_{31}+i\epsilon)^2 + \vec x_{31}^2 } \bigg)^{\Delta/2}\label{circ3}
\eea
and then write the Fourier transform in terms of an integral over a ``loop momentum'' and use the results given in Appendix A for the Fourier transform of each of the three factors composing each one of the four terms in (\ref{circ3}).
The energy step-functions that enter the last three terms are incompatible with the kinematics that we have chosen. That is, with these kinematics, the first term alone accounts for the retarded 3-point function.

As shown in Appendix B, the final answer for the momentum-space retarded 3-point correlator given in (\ref{circ3}) is:
\bea
G_R(p_1,p_2;p_3)&=&\frac{(2\pi)^{8}}{2^{3\Delta-6} \Gamma(\Delta/2)^3\Gamma(3\Delta/2-2)}\delta^4(p_1+p_2+p_3)\int_0^\infty \frac{dz}{z^5}
\times \nonumber\\
&&\bigg[  z^2 (-(E_1-i\epsilon)^2+\vec p_1^2)^{\frac\Delta 2-1} K_{\Delta-2}(z\sqrt{-(E_1-i\epsilon)^2+\vec p_1^2})
\nonumber\\
&& z^2 ( -(E_2-i\epsilon)^2+\vec p_2^2)^{\frac\Delta 2-1} K_{\Delta-2}(z\sqrt{-(E_2-i\epsilon)^2+\vec p_2^2})\nonumber\\
&& z^2 ( -(E_3+i\epsilon))^2+\vec p_3^2)^{\frac\Delta 2-1} K_{\Delta-2}(z\sqrt{-(E_3+i\epsilon)^2+\vec p_3^2})
\bigg].\nn\\
\label{t0ret3pt}
\eea
Each of the three factors present in the previous formula (e.g.
$z^2(-(E_3+i\epsilon)^2+\vec p_3^2)^{\frac\Delta 2-1} \;\;\;\;\;\;\;\;\;$ $K_{\Delta-2}(z
\sqrt{-(E_3+i\epsilon)^2+\vec p_3^2})$) is a  causal (retarded in this case) bulk-to-boundary scalar propagator in the AdS background (see Appendix C). The retarded bulk-to-boundary propagator in AdS can be obtained by analytic continuation, $E\to -i(E+i\epsilon)$, from the Euclidean propagator (\ref{euclt0prop}).

The integral on the right-hand side of (\ref{t0ret3pt}) is not convergent for $\Delta>2$ (the case $\Delta=2$ yields a convergent integral whose closed-form analytic expression  is given in Appendix D). This is dealt with in the usual manner: the integrals could have been regularized using dimensional regularization, as we did before in manipulating the Euclidean 3-point correlator, or one employs boundary cut-off regularization. In this case, the integral over the radial coordinate $z$ is cut-off at $z_B\ll1$, and the propagators are replaced by functions which are normalized at the boundary as in (\ref{normalization}).

\section{Field theory $T{\neq}0$ real-time formalism}
In this section we review a few of the fundamental notions and definitions of the field theory real-time finite-temperature formalism.

In real-time formalism one distinguishes between physical fields
\be
\phi_1(x)=\phi(\vec x, t)
\ee
and doubler fields
\be
\phi_2=\phi(\vec x, t-i\sigma)
\ee
where $\sigma$ is the arbitrary parameter of the Schwinger-Keldysh contour.
Correspondingly, the  real-time n-point Green's functions are defined as
\be
G_{a_1 a_2 \dots a_n}(1,2,\dots n)\equiv(-i)^{n-1} \langle {\cal T}_P \phi_{a_1}(1)
\phi_{a_2}(2)\dots \phi_{a_n}(n)\rangle_\beta, \label{fintempG}
\ee
where $a_1, \dots a_n=1,2$. By ${\cal T}_P$ we denoted the time ordering along the Schwinger-Keldysh contour,  where the $\phi_1(x)$ fields
 are time-ordered, the $\phi_2(x)$ doubler fields
 are anti-time ordered, and lastly with any field $\phi_2$ being defined to have a larger path-time than any other field $\phi_1$. In (\ref{fintempG}) the brackets $\langle\dots\rangle_\beta$ denote the thermal average over all Hamiltonian eigenstates, each being weighted by the Maxwell-Boltzmann factor.
The generating functional for the Green's functions is
\be
Z[J_1,J_2]\equiv\langle {\cal T}_P \exp\bigg( i\int d^4 x(J_1\phi_1-J_2\phi_2)\bigg)\rangle_\beta,
\ee
and so
\be
G_{a_1 a_2\dots a_n}(1,2,\dots n)=i(-1)^{\sum_{i=1}^n a_i}
\frac{\delta^n Z[J_1,J_2]}{\delta J_{a_1}(1)\delta J_{a_2}(2)\dots \delta J_{a_n}(n)} .\label{nptG}
\ee
If $\sigma=\beta/2$, then one can prove that
\be
G_{a_1 a_2\dots a_n}^*=(-1)^{n-1}G_{\bar a_1 \bar a_2\dots \bar a_n}\label{kmsn}
\ee
where $\bar 1=2$ and $\bar 2=1$. These are the Kubo-Martin-Schwinger identities.

Consider now a massless scalar field theory at finite temperature in the real-time formalism. The propagator is a $2\times 2$ matrix
\be
\begin{pmatrix}
D_{11}(p)&D_{12}(p)\\D_{21}(p)&D_{22}(p)
\end{pmatrix}=
\begin{pmatrix}
\frac{i}{-p^2+i\epsilon}+n(|E|)2\pi\delta(p^2)&
(\theta(-E)+n(|E|))2\pi\delta(p^2) e^{\beta E/2}\\
(\theta(E)+n(|E|))2\pi\delta(p^2)e^{-\beta E/2}
&\frac{-i}{-p^2-i\epsilon}+n(|E|)2\pi\delta(p^2)
\end{pmatrix}\label{SK1},
\ee
where
\be
n(|E|)=\frac{1}{e^{\beta|E|}-1},
\ee
and $\beta=1/T$ is the inverse temperature. The off-diagonal components of the Schwinger-Keldysh propagator (\ref{SK1}) are related to the finite-temperature Wightman functions $D^\pm(p)$:
\be
D_{12}=\exp( \beta E/2)D^-,\qquad
D_{21} = \exp(-\beta E/2) D^+ \label{ghostcircling1},
\ee
while the diagonal components are related to the finite-temperature Feynman propagator:
\be
D_{11}=D_F, \qquad \,D_{22}=D_F^*.\label{ghostcircling2}
\ee
In more generality, if the parameter $\sigma$ of the  Schwinger-Keldysh contour is not equal to $\beta/2$ as in the previous formulae, then
\be D_{11}=D_F,\qquad  D_{22}=D_F^*,\qquad  D_{12}(p)=\exp(\sigma E)D^-(p),
\qquad D_{21}(p)=\exp(-\sigma E) D^+(p).
\ee
With the choice $\sigma=\beta/2$, the Schwinger-Keldysh propagator is a symmetric matrix, as in (\ref{SK1}). As it turns out, the finite-temperature computations performed via AdS/CFT following \cite{Herzog:2002pc} yield 2-point correlators with $\sigma=\beta/2$.

The matrix propagator (\ref{SK1}) contains the propagators between all types of vertices: physical field vertices being labeled 1, and doubler field vertices, labeled by 2.  A real-time Feynman $G_{11\dots 1}$ n-point function is computed diagrammatically by summing all distinct diagrams obtained by letting the vertices not connected to external lines be of either type, 1 or 2. The vertices connected to external lines are taken to be of type 1. This turns out to be equivalent to just using the circling rules, and the $D_F, D_F^*, D^\pm$ propagators between circled/uncircled vertices, with only the internal vertices being of either circled/uncircled type \cite{Kobes:1985kc}.
Moreover, the largest time equation of Veltman still holds at finite temperature.

These observations led Kobes \cite{kobes} to define a ``causal'' n-point function, by adding all diagrams with a vertex circled or uncircled, with the exception of the vertex associated with ``the largest time'', which remains uncircled. This is very much the same prescription used at zero temperature. It is this retarded n-point function which is obtained by the analytic continuation of the n-point function computed in imaginary-time formalism \cite{kobes}\footnote{The other real-time finite-temperature correlators do not enjoy such a simple relationship with the imaginary-time correlator.}.

For example, a
causal/retarded 3-point function, 
 with the outgoing momentum $r=-(p+q)$ vertex corresponding to ``the largest time'' is equal to
\be
G_R(q,p;r)=G_{111}-e^{-\beta E_p/2}G_{121}-e^{-\beta E_q/2}G_{211}+e^{\beta E_r/2}G_{221}\label{causal3point}.
\ee
In addition to the identities which follow from (\ref{kmsn})
\be
G_{111}=G_{222}^*,\; G_{121}=G_{212}^*,\;
G_{112}=G_{221}^*,\; G_{211}=G_{122}^*,\label{kms3ft}
\ee
the largest time equation yields one more identity
\be
0=G_{111}-e^{-\beta E_p/2}G_{121}-e^{-\beta E_q/2}G_{211}+e^{\beta E_r/2}G_{221}+e^{\beta E_p/2}G_{212}+e^{\beta E_q/2}G_{122}-e^{-\beta E_r/2}G_{112}-G_{222}\label{largesttime},
\ee
which, using (\ref{kms3ft}), can be re-written as
\bea
\!\!\!\!\!\!\!\!\!\!\!\!\!\!\!\!\!&&\!\!\!\!\!\!\!\!-\sinh(\omega_r \pi)\Re G_{112}(q,p) = \sinh(\omega_q \pi)\Re G_{211}(q,p)+\sinh(\omega_p \pi) \Re G_{121}(q,p),\nn\\
\!\!\!\!\!\!\!\!\!\!\!\!\!\!\!\!\!&&\!\!\!\!\!\!\!\!-\Im G_{111}(q,p)+\cosh(\omega_p\pi)\Im G_{121}(q,p)+\cosh(\omega_q \pi)\Im G_{211}(q,p)+\cosh(\omega_r \pi)\Im G_{112}(q,p)=0.\nonumber\\
\label{largest3}
\eea
Substituting (\ref{largest3}) into (\ref{causal3point}) leads to a simpler expression for the causal 3-point function \cite{kobes}:
\bea
&&\Re G_R(q,p;r)=\Re G_{111} + \frac{\sinh(\beta E_q/2)}{\sinh(\beta E_r/2)}
\Re G_{121}+\frac{\sinh(\beta E_p/2)}{\sinh(\beta E_r/2)}\Re G_{211},
\\
&&\Im G_R(q,p;r)=-\tanh(\beta E_r/2)\bigg(\Im G_{111}+\frac{\sinh(\beta E_q/2)}{\sinh(\beta E_r/2)}\Im G_{121}+\frac{\sinh(\beta E_p/2)}{\sinh(\beta E_r/2)}\Im G_{211}\bigg).
\nonumber\\
\eea

\section{Real-time  AdS-S propagators and Circling rules}

Some of the results presented in this section are known in the literature \cite{CaronHuot:2009iq}. However we cound not find references giving a comprehensive picture of the material included here. Also since the literature is not conclusive in terms of how to best formulate real-time AdS/CFT \cite{Balasubramanian:1998sn, Balasubramanian:1998de, Giddings:1999jq, Son:2002sd, Herzog:2002pc, Skenderis:2008dh}, we have decided to spend some time discussing the various bulk-to-boundary propagators (retarded, Feynman, etc...) in AdS-S. These propagators will consitute the basis of a diagrammatic expansion in terms of tree-level gravity diagrams, modeled on the finite-temperature real-time field theory formalism. The AdS-S retarded propagator is the one which behaves like an incoming wave at the horizon. We discuss its analytic properties, causal nature and zero-temperature limit. The Feynman propagator which corresponds to the thermal ``Kruskal'' vacuum is obtained from the retarded propagator using (\ref{fdt}), which is typical of finite-temperature systems. Next we formulate the AdS-S circling rules as the curved space counterpart of the same rules introduced in the previous sections. This will help us develop a better perspective on the prescription for computing the real-time finite-temperature correlators in Section 5.

\subsection{The AdS-Schwarzschild Geometry}

The Anti de Sitter-Schwarzschild (AdS-S) black hole (times a five-dimenional sphere $S^5$) is the holographic dual of the finite-temperature ${\cal N}{=}4$ super Yang-Mills theory with gauge group $SU(N_c)$, in the limit $N_c\gg1$, and in the deconfined (high-temperature) phase \cite{Witten:1998zw}.

The AdS-S metric is usually written as
\bea
&&ds_{10}^2 = \frac{r^2}{R^2}(-f(r) dt^2 + d\vec x^2)+\frac{R^2}{r^2 f(r)}dr^2+R^2 d\Omega_5 \nonumber\\
&&\;\;\;\;\;\;\;\;=R^2 \bigg(\frac{-f(z) dt^2 + d\vec x^2 + \frac{dz^2}{ f(z)}}{z^2} + d\Omega_5^2\bigg), \qquad z=\frac {R^2}{r},\label{metric2}\\
&&f(r)= 1-\frac{r_0^4}{r^4}\label{blackening},
\eea
where the position of the black hole horizon is at $r_0$ and the asymptotic region of the black hole geometry is at $r=\infty$ (or $z=0$).
The black hole Hawking temperature is
\be
T_H=\frac{r_0}{\pi R^2}.
\ee
A better suited choice of coordinates is
\bea
&&u=\frac{r_0^2}{R^4} z^2, \qquad f(u)= 1-u^2,
\nonumber\\
&&ds_5^2=\frac{r_0^2}{R^2}\bigg(-\frac{1-u^2}{u}dt^2+\frac{d\vec x^2}{u}\bigg)
+\frac{R^2}{4(1-u^2)u^2} du^2\nonumber\\
&&\;\;\;\;\;\;\;=\frac{\pi^2 T_H^2 R^2}{u}(-(1-u^2)dt^2+d\vec x^2)+\frac{R^2}{4(1-u^2) u^2}du^2\label{adss}.
\eea
In these new coordinates, the horizon is at $u=1$, the singularity is at $u=\infty$ and the asymptotic region is at $u=0$.

It has been conjectured that finite-temperature correlators should be related via AdS/CFT to $n$-point functions of supergravity living in the maximally extended AdS-S geometry \cite{Herzog:2002pc}. Therefore, we will begin by switching to global coordinates which describe the extended geometry. We will look for coordinates $t_K$, $x_K$ which are the analog of Kruskal coordinates for Schwarzschild black holes in flat space. These can be found by starting with the ansatz:
\be
t_K={\cal R}(u)\sinh\left({2\pi T_H}\,t\right),\qquad x_K={\cal R}(u)\cosh\left({2\pi T_H}\,t\right).\label{kruskal}
\ee
In terms of $t_K$ and $x_K$, we demand that the metric have the form
\be
ds^2=W(u)\left[-dt_K^2+dx_K^2\right]+{(\pi T_HR)^2\over u}d\vec{x}^2+R^2d\Omega_5^2.
\ee
Matching this metric to our black hole metric requires
\bea
&&{\cal R}(u)=\exp\left[-\tan^{-1}\sqrt{u}-{1\over2}\log\left(1+\sqrt{u}\over1-\sqrt{u}\right)\right],\\ &&W(u)={f(u)\over 4u{\cal R}^2(u)}.
\eea
Since $x_K^2-t_K^2={\cal R}(u)^2$,  the event horizon is described by two intersecting lines in the $t_K/x_K$ plane:
\be
\hbox{event horizon:}\qquad x_K^2-t_K^2={\cal R}^2(1)=0.\label{Khorizon}
\ee
In the above expression for ${\cal R}(u)$, we have also taken the liberty of setting to zero an integration constant which would appear as an overall rescaling of ${\cal R}$. This fixes the distance between the origin $(t_K,x_K)=(0,0)$ and the $AdS$ boundary to be 1:
\be
\hbox{AdS boundary:}\qquad x_K^2-t_K^2={\cal R}(0)^2=1.\label{Kboundary}
\ee
This choice also fixes the distance between the origin and the black hole singularity:
\be
\hbox{singularity:}\qquad x_K^2-t_K^2={\cal R}^2(\infty)=-e^{-\pi}.\label{Ksingularity}
\ee
It is apparent from the above expressions for the event horizon, AdS boundary and black hole singularity that, in terms of $t_K$ and $x_K$, there are two event horizons, two AdS boundaries, and two singularities, if $x_K$ is analytically continued to negative values. There is no obstacle to performing this analytic continuation, and the result is the extended AdS-S geometry. In the extended geometry, $t_K$ and $x_K$ have the ranges $t_K\in(-\infty,\infty)$ and $x_K\in(-\infty,\infty)$.The Penrose diagram of the AdS-S geometry is given in Fig.2\footnote{We took the artistic license to draw a square Penrose diagram, but this is accurate only for the 3-dimensional AdS-S black hole.}. The fact that there are  two AdS boundaries is particularly important for the computation of time-dependent correlators.
\begin{figure}
\begin{center}
\includegraphics[width=1.3in]{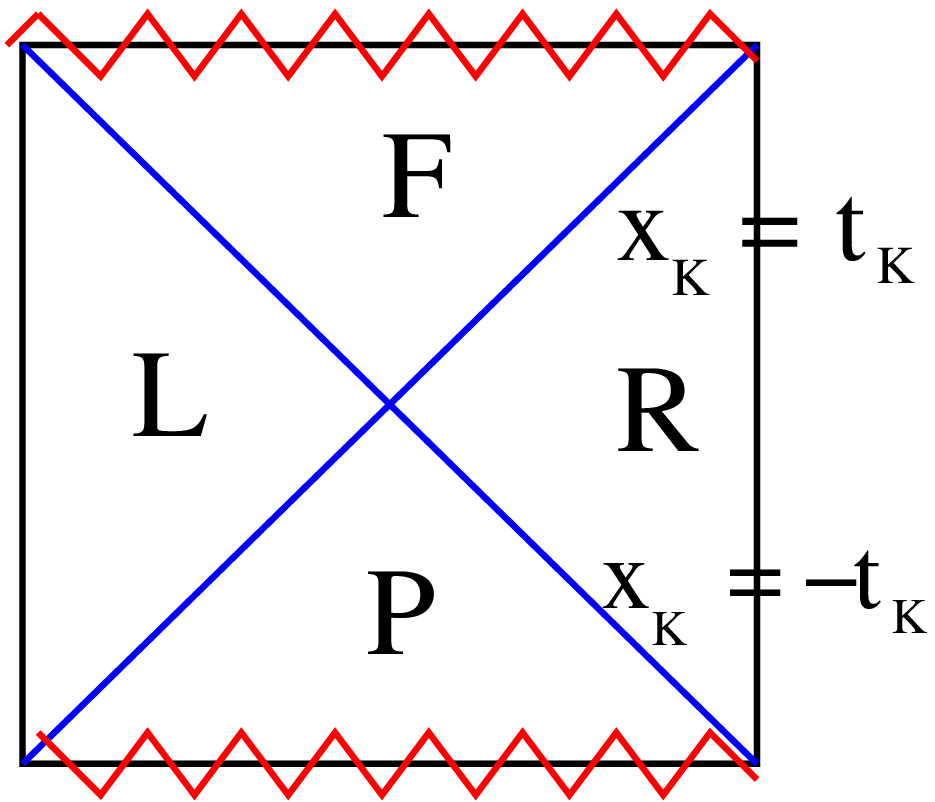} \\
Figure 2: The AdS-S Penrose diagram
\end{center}
\end{figure}

In what follows we will ignore the fluctuations in the compact $S^5$ directions, and consider only gravity fluctuations in the 5-dimensional AdS-Scharzschild geometry (\ref{adss}).
\subsection{The retarded bulk-to-boundary scalar propagator in AdS-S}

\subsubsection{Minkowski signature}

Consider a minimally coupled massless scalar propagating in the black hole background.
To terms quadratic in fields, its equation of motion is
\be
\partial_\mu(\sqrt{-g} g^{\mu\nu} \partial_\nu \phi(x^\mu,u))=0.
\ee
After Fourier-transforming along the $x^\mu$ coordinates to momentum space, 
the scalar field equation of motion becomes
\be
F''-\frac{1+u^2}{u(1-u^2)}F'+\bigg(\frac{\omega^2}{u(1-u^2)^2}
-\frac{|\vec p|^2}{u(1-u^2)}\bigg)F=0 \label{scalar_eom},
\ee
where $\phi(p^\mu,u)=F(\omega, \vec p,u)\phi_0(p^\mu)$, $\phi_0(p^\mu)$ is the boundary value of the scalar field, and prime denotes differentiation with respect to the bulk coordinate $u$. The variables $\omega, \vec p$ are dimensionless quantities defined as
\be
\omega= \frac{E}{2\pi T_H},\vec p=\frac{\vec P}{2\pi T_H},
\ee
where $E, \vec P$ are the energy and momentum associated with the scalar field modes. The norm of the dimensionless spatial momentum is denoted by $|\vec p|$.
The bulk-to-boundary propagator is further normalized to 1 at the boundary: $F(u=0)=1$. The 5-dimensional AdS-S bulk-to-boundary propagators admit analytic expressions in terms of Heun's functions (see also \cite{Son:2002sd}, \cite{Starinets:2002br}).

We  begin by making the substitution
\be
F=(1-u)^{\frac{-i\omega}2}(1+u)^{\frac{\omega}2}H.
\ee
Then, $H$ obeys the differential equation
\be
H''+
\bigg(-\frac 1u+\frac{-1+i\omega}{1-u}+\frac{1+\omega}{1+u}\bigg)H'
+\frac{2(\omega^2-|\vec{p}|^2)-(1+i)\omega+i\omega^2u}{2u(1-u^2)}H=0,\label{heun1}
\ee
which is of Heun type (see Appendix E), with parameters\footnote{The bulk-to-boundary propagator for a massive scalar field can also be expressed in terms of a Heun's function. In this case, one starts with the ansatz $F=u^\lambda(1-u)^{\frac{-i\omega}2}(1+u)^{\frac{\omega}2}H$, where $\lambda=\frac{\Delta}{2}$ or $2-\frac{\Delta}{2}$. The Heun parameters will then also depend on the conformal dimension $\Delta$.}
\be
d=-1,\;q=\omega^2-|\vec p|^2-\frac{1+i}2\omega,\;\alpha=\beta=\frac{1-i}{2}\omega,\;\gamma=-1,\;\delta=1-i\omega.
\ee
The two independent solutions of (\ref{heun1}) are
\bea
&&H_1(u)\equiv \lim_{\gamma\to -1}\bigg[Hl(-1,q,\alpha,\beta,\gamma,\delta ;u),\nonumber\\
&&\;\;\;\;\;\;\;\;+\frac{1}{2}\frac{(\omega^2-|\vec p|^2)^2}{\gamma+1}u^{1-\gamma}Hl(-1,q',\alpha',\beta',\gamma',\delta;u)\bigg]\label{h1}\\
&&q' = \omega^2-|\vec p|^2-\frac{1+i}2\omega-(\gamma-1)(\omega(1+i)-\gamma-1)\nonumber\\
&&\alpha'=\beta'=\frac{1-i}2\omega-\gamma+1,\gamma'=2-\gamma\nonumber\\
&&{\rm {and}},\nonumber\\
&&H_2(u)\equiv u^2 Hl(-1,q',\alpha',\beta',\gamma',\delta;u),\label{h2}
\eea
where in the definition of $H_1$  $\gamma$ is kept arbitrary prior to taking the limit (see equation (E.5) and related discussion in Appendix E). In $H_2$, $\gamma$ can be set to $-1$ directly.
For future reference we give the small $u$ expansions of $H_1(u)$ and $H_2(u)$:
\bea
H_1(u)&=&1+ c_1 u + (c_2\ln(u)+c_2') u^2 +\dots\nn\\
H_2(u)&=& u^2+ \dots \label{smalluheun}
\eea
where 
\bea
&&c_1=q=-{1+i\over2}\omega+\omega^2-|\vec{p}|^2,\nn\\&& c_2=-{(\omega^2-|\vec{p}|^2)^2\over2},\nn\\
&&c_2'=\frac{q}2(1-q-\omega (1+i)).
\eea

Next, we focus on constructing the solution of (\ref{heun1}) which corresponds to a purely incoming wave at the horizon. This is the retarded bulk-to-boundary propagator in the black hole background \cite{CaronHuot:2009iq}:
\be
F(u=1)={\rm incoming\; wave}\qquad {\rm and} \qquad F(u=0)=1\footnote{For the case under consideration, that of a minimally coupled {\it massless} scalar,  the bulk-to-boundary propagator remains finite at $u=0$ and therefore can be normalized at the boundary. However, the 3-point correlator is expressed as an integral over the radial coordinate $u$ and has a  divergent integrand at $u=0$ because of the measure factor. The integral is regularized usually by  boundary cut-off regularization, in which case the bulk-to-boundary propagators are normalized at $u_B\ll 1$. }.
\ee
An incoming wave at the horizon ($u=1$) behaves as $(1-u)^{\frac{-i\omega}2}$, whereas the outgoing wave is $(1-u)^{\frac{i\omega}2}$.

To this end, we make a change of variable,
\be
w=1-u, \qquad H(u)=\tilde H(w).
\ee
After substituting into (\ref{heun1}), we arrive at another Heun equation for $\tilde H$. The incoming wave at the horizon corresponds to the solution
\be
H_3(u)\equiv  Hl (2, \omega^2(-1-\frac{i}2)+\frac{1+i}2\omega + |\vec p|^2,\frac{1-i}2\omega,\frac{1-i}2\omega, 1-i\omega,-1;1-u).
\ee
Finally, the bulk-to-boundary propagator corresponding to an incoming wave can be expressed as
\be
F=B (1-u)^{-\frac{i\omega}2} (1+u)^{\frac \omega 2} H_3(u)\label{h3},
\ee
where $B$ is a normalization coefficient
\be
B=\frac{1}{H_3(0)}.
\ee
Alternatively, we can write
\be
F= (1-u)^{-\frac{i\omega}2} (1+u)^{\frac \omega 2}\bigg(H_1(u) + A H_2(u)\bigg),
\label{F2}\ee
where the coefficient $A$ is\footnote{In practice, in numerical computations we have matched (\ref{h3}) and (\ref{F2}) at some value $u_m=0.5$ where both (\ref{h3}) and (\ref{F2}) are within the radius of convergence of the corresponding Heun's functions.}
\be
A=\bigg(\frac{1}{H_3(0)}-H_1(1)\bigg)\frac{1}{H_2(1)}.
\ee
$A$ can be solved for in the limit of small $\omega$ and $|\vec{p}|^2$ directly from (\ref{heun1}), by writing $H(u)=1-{\cal H}  h(u)\equiv 1-(\frac{1+i}2\omega+ |\vec p|^2)/2 \,h(u)$ and solving for $h(u)$ such that to order ${\cal H}$ the solution one finds is regular at the horizon and it is normalized to 1 at the boundary. This allows computing the terms linear and quadratic in $u$, to order ${\cal H}$. Recalling that the terms quadratic in $u$ in $H$ have a coefficient $c_2'+A$, we get 
\be
A={1+i\over2}\omega +{|\vec{p}|^2}+O(\omega^2,\omega |\vec{p}|^2,|\vec{p}|^4).\label{A small omega limit}
\ee

Earlier we have advertised that $F$ is the retarded bulk-to-boundary propagator.
To better understand that is so, we will show that $F$ is obtained by analytic continuation of the Euclidean signature propagator which is regular at the origin (i.e. at $u=1$) for values of the Euclidean frequency such that $\Re(\omega_E)>0$. Then we will argue that $F$ is analytic in the upper half $\omega$-plane, and its Fourier transform to position space is a causal function, with support inside the future light cone. Lastly, we will show that the zero-temperature limit of the AdS-S retarded propagator is, as expected, another causal propagator,  the AdS retarded bulk-to-boundary propagator.

\subsubsection{Euclidean signature}

Here we address the same problem of a bulk-to-boundary propagator, but in the Euclidean version of the AdS-S black hole:
\be
ds^2_5={(\pi T_HR)^2\over u}\left[(1-u^2)d t_E^2+d\vec{x}^2\right]+{R^2\over4u^2(1-u^2)}du^2. \label{emetric}
\ee
The metric in (\ref{emetric}) has a conical singularity at $u=1$, unless the Euclidean time $t_E$ is periodic with period $1/T_H$. Then, the origin is a regular point, which corresponds to the black hole horizon in Minkowski signature. Also in the Euclideanized geometry, there is only a single (Euclidean) AdS boundary. One way to see this is to return to our earlier analysis of Kruskal coordinates. We could imagine writing a Euclidean analog of equation (\ref{kruskal}):
\be
t_K={\cal R}(u)\sin\left({2\pi T_H}\,t_E\right), \qquad x_K={\cal R}(u)\cos\left({2\pi T_H}\,t_E\right).
\ee
In terms of these coordinates, we no longer get two disconnected boundaries and two horizons. Instead, the horizon is just a point ($t_K=x_K=0$), and there is a single connected boundary which is just a circle of radius one: $t_K^2+x_K^2=1$. The black hole singularity is not part of the Euclidean geometry.

The wave equation for a massless scalar in this geometry is
\be
\phi''-{1+u^2\over u(1-u^2)}\phi'-{1\over u(1-u^2)^2}(\omega_E^2+(1-u^2)|\vec{k}|^2)\phi=0.
\ee
We are interested in positive frequency solutions that are regular at the origin ($u=1$), so we set
\be
F_E(u)\equiv{\phi(u,k)\over \phi_0(k)}=(1-u)^{\omega_E/2}(1+u)^{i\omega_E/2}H(u,k)\label{Fe}
\ee
where the dimensionless Euclidean 4-vector $k^\mu$ is $k^\mu=(\omega_E,\vec k)$.
$\phi_0(k)$ is the value of the scalar field on the boundary of Euclidean AdS-S. Given the periodicity of $t_E$, we must conclude that $\omega_E$ is integer-valued. On the other hand, for the purpose of performing analytic continuation, we will allow $\omega_E$ to become complex-valued.

Plugging this ansatz into the equation for $\phi$, we find that $H$ satisfies the Heun equation:
\be
H''+\left({\gamma\over u}+{\delta\over u-1}+{\epsilon\over u-d}\right)H'+{\alpha\beta u-q\over u(u-1)(u-d)}H=0,
\ee
with the various parameters given by
\bea
&&\gamma=-1,\quad \delta=1+\omega_E,\quad \epsilon=1+i\omega_E,\quad \alpha=\beta={1+i\over2}\omega_E,\nn\\ &&q={1-i\over2}\omega_E-\omega_E^2-|\vec{k}|^2,\quad d=-1.\label{Heunparameters}
\eea
It is easy to check that the condition $\epsilon=\alpha+\beta-\gamma-\delta+1$ is satisfied.
As in the previous section, we can write the solution to the Heun equation either in terms of Heun's functions defined in a neighborhood of the boundary $u=0$ or in terms of functions defined in a neighborhood of the horizon $u=1$. In the former case, we have
\be
F_E=(1-u)^{\omega_E/2}(1+u)^{i\omega_E/2}
(H_1(u)+A H_2(u))
\label{HeunEucl},
\ee
where we normalized the Euclidean propagator 
$u=0$.  $H_1$ and $H_2$ are two independent solutions to the Heun equation which are defined as in (\ref{h1}) and (\ref{h2}), and whose parameters are given in (\ref{Heunparameters}).

Alternatively, we may express $F_E$ in terms of a Heun's function defined near $u=1$\footnote{
The solution $F_E$ which is regular at the origin for negative frequencies is $$\frac
{(1-u)^{-\omega_E/2}(1+u)^{i\omega_E/2}Hl(1-d,-q+(\delta-1)\gamma d+(\alpha-\delta+1)(\beta-\delta+1);\beta-\delta+1,\alpha-\delta+1,2-\delta,\gamma;1-u)}{Hl(1-d,-q+(\delta-1)\gamma d+(\alpha-\delta+1)(\beta-\delta+1);\beta-\delta+1,\alpha-\delta+1,2-\delta,\gamma;1)}
.$$ The analytic continuation of this Euclidean solution is the advanced propagator, which in momentum space is the complex conjugate of the retarded propagator $F$.}:
\bea
F_E(u)&=&{(1-u)^{\omega_E/2}(1+u)^{i\omega_E/2}Hl(1-d,\alpha\beta-q;\alpha,\beta,\delta,\gamma;1-u)\over Hl(1-d,\alpha\beta-q;\alpha,\beta,\delta,\gamma;1)}.\label{fnorm}\nn\\
&\equiv&\frac{(1-u)^{\omega_E/2}(1+u)^{i\omega_E/2}H_3(u)}{H_3(0)}
\eea
By matching these two expressions for $F_E$, as in the previous section, we have
\be
A= {1\over H_2(1)}\left[{1\over H_3(0)}-H_1(1)\right].
\ee
Using (\ref{fnorm}) we have generated plots for a wide range of the parameters
$\omega_E,\vec k^2$ and $u$. We are only looking at the right half $\omega_E$ complex plane since the solution (\ref{fnorm}) gives a regular Euclidean propagator only for $\Re(\omega_E)\geq 0$. From these plots (a couple of which are being shown in Figure 3) we have determined that $F_E$ is a smooth function with no infinite singularities for $\Re(\omega_E)>0$. This is the expected behavior of an analytic function of $\omega_E$ in this region of the complex plane.
\begin{figure}[h]
\centering
\includegraphics[width=7 cm]{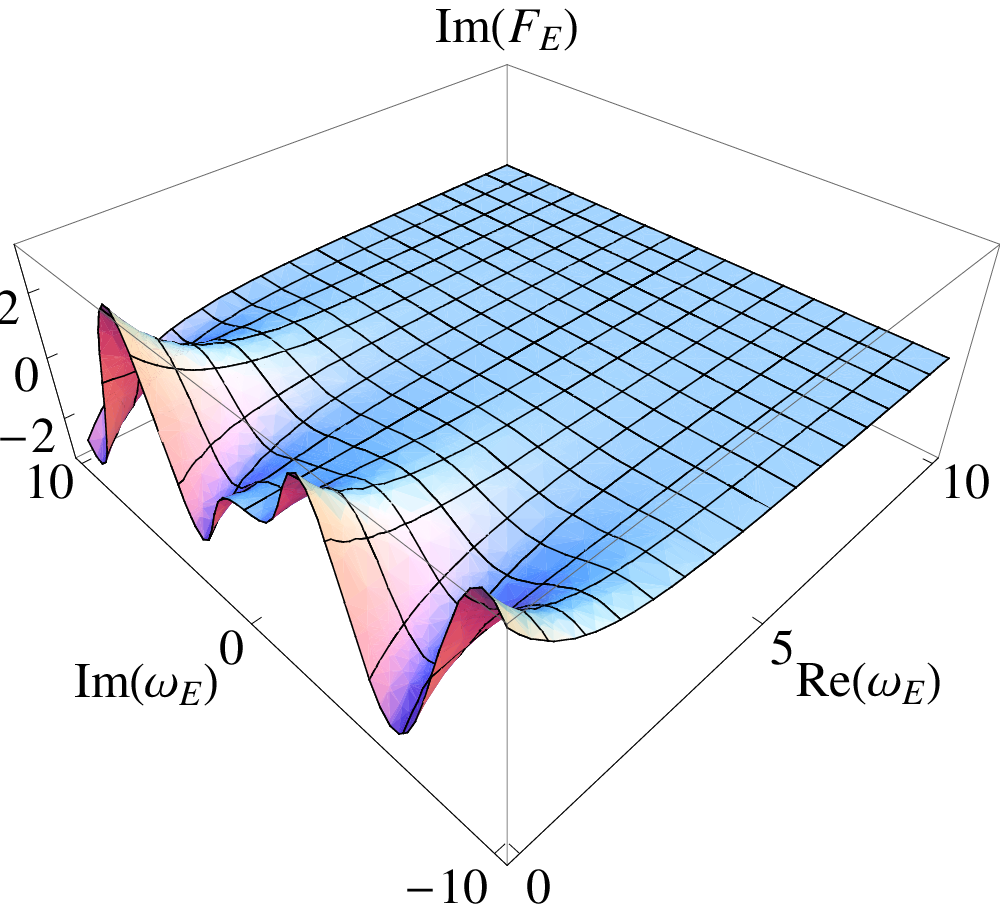}\qquad
\includegraphics[width=7 cm]{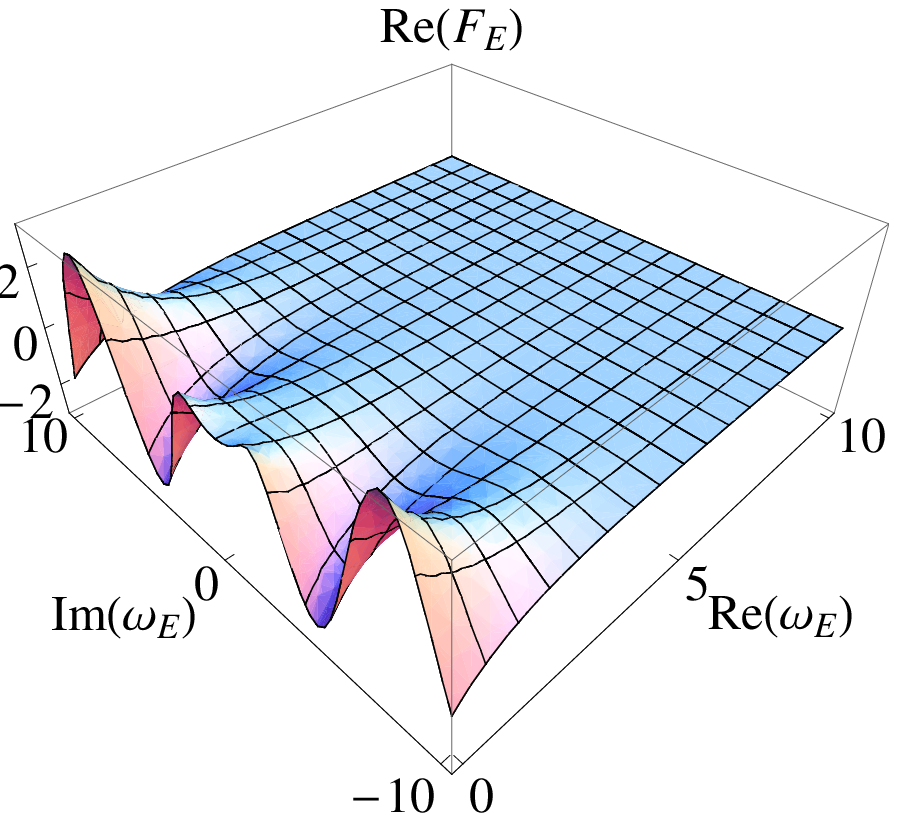}\\
Figure 3: Re and Im parts of the Euclidean bulk-to-boundary propagator $F_E$, with $u=0.5$ and $\vec k^2=1$.
\end{figure}
The retarded propagator which is obtained from the Euclidean propagator by the usual analytic continuation,
\be
F(\omega,\vec p, u) = F_E(-i(\omega+i\epsilon)),\vec p, u),
\ee
enjoys the same properties (smooth, free of infinite singularities) in the upper half Minkowski frequency plane $\Im(\omega)>0$.

\subsubsection{Causality}

We will next argue that the Fourier transform of the finite-temperature retarded propagator, $F(\omega,\vec p,u)$, is causal, i.e. has support inside the forward light cone.  This is closely related to the analytic properties of $F$.
Our analysis will be restricted for simplicity to the case when two spacetime points are separated in $t$ and $u$ but have the same coordinate $\vec x=0$.

First, we construct the null AdS-S geodesics at $\vec x=0$:
\be
0=-\frac{1-u^2}{u}\frac{r_0^2}{R^2} dt^2 + \frac{R^2}{4(1-u^2)u^2}du^2
\ee
with solution
\be
t(u)=\frac{1}{2\pi T_H}\bigg[{\rm arctan}(\sqrt{u})+{1\over2}\log\left(1+\sqrt{u}\over 1-\sqrt{u}\right)\bigg]\equiv \frac{1}{2\pi T_H} \tau(u).
\ee
Values of $t$ lying outside the forward light cone obey the condition $t<t(u)$.

Next we infer the behaviour of $F(\omega,\vec{p},u)$ at large frequency ($\omega\gg 1$) and fixed $u$\footnote{ This can be done in the WKB approximation. The proportionality coefficient is fixed by solving $F(u)$ for small $u$, such that it is normalized to 1 at $u=0$, and constructing the interpolating function: $i\pi (\omega + i\epsilon)^2 u \sqrt{\tau(u)}/\sqrt{2\sqrt u}\,H_2^{(1)}((\omega+i\epsilon) \tau(u))$.  }:
\be
F= \sqrt{\pi} (\omega+i\epsilon)^{3/2} u^{3/4} e^{-\frac{3i\pi}4} e^{i\omega\tau(u)}+\dots \label{1up},
\ee
Plots of the exact $F$ along with this leading order behavior are shown in Fig. 4. When Fourier-transforming to position space, the leading term in (\ref{1up}) yields
\be
-\frac{3\sqrt{2\pi}u^{3/4}}{8}\theta(t-t(u))\frac{1}{(2\pi T_H\,t -\tau(u))^{5/2}},
\ee
 with support inside the forward light cone. The next subleading term in (\ref{1up}) behaves like $(\omega+i\epsilon)^{1/2}e^{i\omega \tau(u)}$. Its Fourier transform also has support inside the forward light cone.
 In order to analyze the behavior of the remaining contributions to $F$, we consider the following integral:
\be
{1\over2\pi}\int_{-\infty}^\infty d\omega e^{-i\omega t}\left[F(\omega,\vec{p},u)-\bigg(\sqrt{\pi}(\omega+i\epsilon)^{3/2} u^{3/4}e^{-\frac{3i\pi}4} - (\omega+i\epsilon)^{1/2}f(\omega,\vec{p},u)\bigg)e^{i\omega\tau(u)}\right],
\ee
where $f(\omega, \vec p, u)$ refers to the first sub-leading term in (\ref{1up}). 
For those values of $t$ which lie outside the light cone ($t(u)-t>0$), we may close the contour in the upper half $\omega$ plane. The integrand has no poles or cuts in this region, and so the integral vanishes. Since the retarded propagator is causal for $\vec x=0$, and since the zero-temperature limit of the retarded propagator is also causal (see Section 4.2.4 and Appendix C), this provides strong evidence that $F(\omega, \vec p, u)$ is causal in general.
 
\begin{figure}
\begin{center}
\includegraphics[width=2.7in]{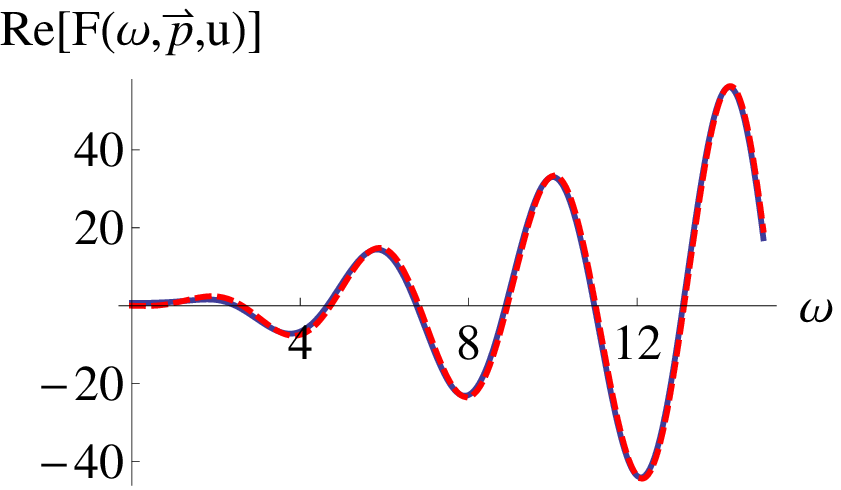}\;\;\;\;\;\;\;\;
\includegraphics[width=2.7in]{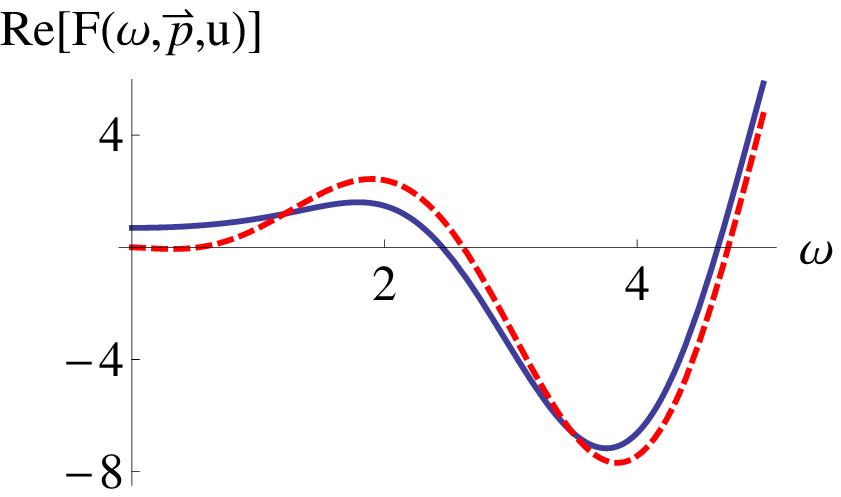} \\
Figure 4: The exact $F(\omega,\vec{p},u)$ (blue, solid) and its leading-order behavior (\ref{1up}) in the large $\omega$ limit (red, dashed) with $u=0.5$ and $|\vec{p}|=1$. The plot on the right is a zoom at small values of $\omega$.
\end{center}
\end{figure}

\subsubsection{Zero-temperature limit}
As we move farther away from the horizon, we expect that near the boundary, the retarded propagator $F$ approaches the AdS retarded propagator (which is expressed in terms of a Hankel function) for $|\omega^2-\vec p^2|$ sufficiently large. To simplify our discussion of the small $u$ limit we set $\vec p=0$.  In this limit, and switching to Euclidean signature, the equation of motion for the massless scalar $\phi$ is
\be
\phi_E''-{1\over u}\phi_E'-{\omega_E^2\over u}\phi_E=0.
\ee
The general solution to this equation is
\be
\phi_E=c_1 u K_2(2\sqrt{u}\omega_E)+c_2 u I_2(2\sqrt{u}\omega_E).\label{above0}
\ee
This is, of course, the general solution to the scalar wave equation in pure AdS, i.e. the zero-temperature limit of AdS-S.
The $T_H$ dependence in (\ref{above0}) actually drops out, since $\omega_E\sqrt{u}=Ez/2$.

By allowing the range of $z=\sqrt{u}/(\pi T_H)$ to extend from 0 to infinity, keeping the Bessel function which is well behaved in the interior, 
\be
F_E(u,\omega_E)=2u\omega_E^2 K_2(2\sqrt{u}\omega_E),\qquad u\ll1 ,
\ee
and analytically continuing this result to Minkowski space $(\omega_E\to -i(\omega+i\epsilon))$, we obtain
\be
F(u,\omega)=i\pi u\omega^2H_2^{(1)}(2\sqrt{u}\omega), \qquad u\ll1,
\ee
where $H_2^{(1)}$ is a Hankel function of the first kind. In Figure 5 we plotted both the retarded propagator $F$ and the zero-temperature propagator at $u=0.1$, with the two curves almost perfectly overlapping (the two propagators differ for small values of $\omega$). For comparison, we also plotted the finite and zero-temperature propagator for $u=0.5$.
\begin{figure}
\begin{center}
\includegraphics[width=2.7in]{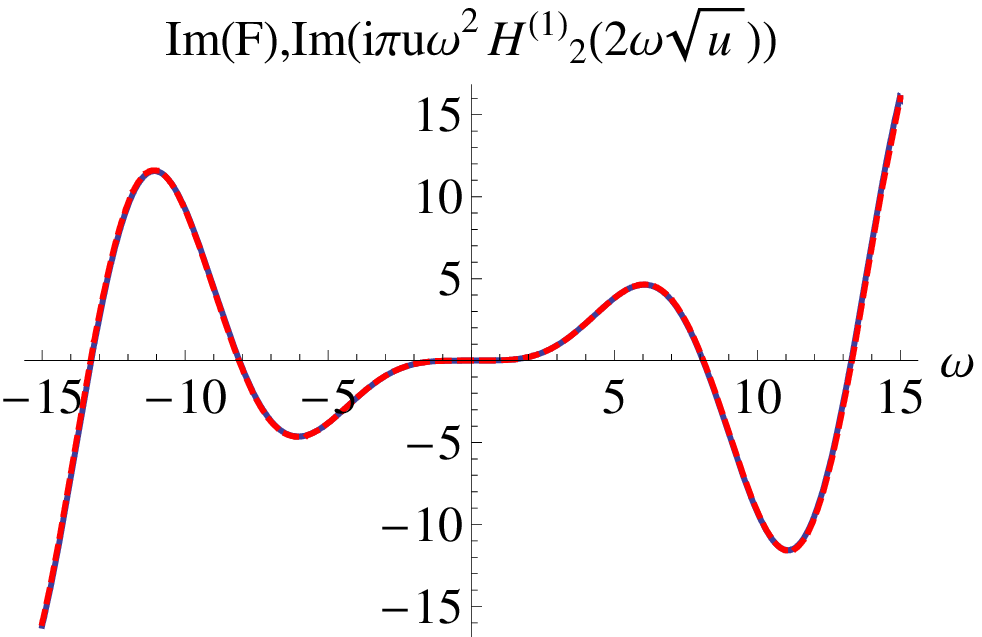} \;\;\;\;\;\;\;\;\;\;
\includegraphics[width=2.7in]{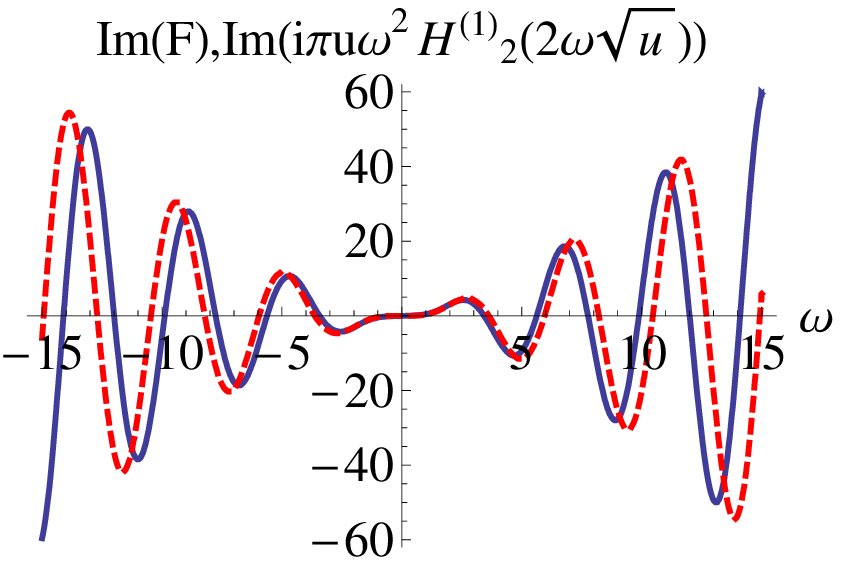}\\
Figure 5: The zero (in red, dashed) and finite-temperature (in blue, solid) bulk-to-boundary propagators at u=0.1 (left) and at u=0.5 (right), with $\vec p=0$.
\end{center}
\end{figure}

\subsection{Thermal Feynman propagator and AdS-S circling rules}
Gibbons and Perry \cite{Gibbons:1976es} noticed that the Hartle-Hawking definition of the Feynman propagator in a black hole background corresponds to the ``Kruskal'' vacuum. In other words it is periodic in imaginary time, and therefore can be identified with a thermal Green's function.

To better understand this, we turn to the Rindler background example. Starting from 2-dimensional Minkowski space, $ds^2=-dt^2+dx^2$ with the identifications
$x=\exp(\rho)\cosh(\tau), t=\exp(\rho)\sinh(\tau)$ one finds the 2-dimensional Rindler space metric $ds^2=\exp(2\rho)\,(d\rho^2-d\tau^2)$. The Feynman propagator corresponding to the Rindler vacuum is proportional to $\ln((\Delta\rho+\Delta\tau)(\Delta\rho-\Delta\tau))$. On the other hand, the Feynman propagator corresponding to the Minkowski vacuum is proportional to $\ln((\Delta x+\Delta t)(\Delta x-\Delta t))$. When writing the Minkowski vacuum Feynman propagator in Rindler coordinates, one discovers that it is periodic in imaginary Rindler time: $\ln(\cosh(\tau)-\cosh(\rho))+f(\rho)$. We arrive at the same answer when starting from the Rindler space retarded propagator in momentum space, in conjunction with the relation
\be
G_F(E,P)=\Re G_R(E, P) + i\coth(\beta E/2) \Im G_R(E,P)  ,\label{use}
\ee
where $\beta=1/T=2\pi$. Lastly, we recall that the 2-dimensional Feynman propagator computed at finite temperature in imaginary-time formalism is $\ln(\cos(\tau_E)-\cosh(\rho))$. Thus the Feynman propagator derived from  (\ref{use}) is a thermal Green's function by construction.

Based on these observations, we proceed to define the Feynman Green's function corresponding to the ``Kruskal'' vacuum of the AdS-S black hole by
\be
{\cal G}_F(\omega,\vec p, u)=\Re {\cal G}_R(\omega,\vec p, u)+ i \coth(\omega\pi) \Im {\cal G}_R(\omega,\vec p, u)\label{th}
\ee
where we used the fact that $\omega=E/(2\pi T_H)$.
Since the retarded Green's function is analytic in the upper half-plane, (\ref{th}) ensures that ${\cal G}_F$ will be a thermal Green's function, periodic in imaginary time.
At this stage we can compute the other Green's functions by the usual relations:
\be
{\cal G}^-={\cal G}_F-{\cal G}_R, \qquad  {\cal G}^+= {\cal G}_F-{\cal G}_A.
\ee
We can represent these bulk-to-boundary propagators using the same circling rules as before: the bulk-to-boundary propagator between circled and uncircled vertices is ${\cal D}^-$, between uncircled and circled vertices is ${\cal D}^+$, and the propagator between two uncircled (or two circled) vertices is ${\cal D}_F$ (or ${\cal D}_F^*$, respectively).

\section{Real-time finite-temperature AdS/CFT correlators}
\subsection{The scalar two-point function}
In this section we review briefly the scalar 2-point function computation in real time and at finite temperature, following Son and Herzog \cite{Herzog:2002pc}. We have already discussed the difference between Euclidean and Minkowski signature AdS-S geometries. We are interested in clarifying the issue of what bulk region must the tree-level gravity diagrams be  integrated over.
\subsubsection{Minkowski Signature: the Schwinger-Keldysh Propagator}
From the work of Son and Herzog \cite{Herzog:2002pc} we know that in real time, the gravity fields need to be specified on both time-like boundaries of the Penrose diagram. Therefore, one expects that one needs to integrate over the position of the bulk vertices over at least the R and L quadrants of the Penrose diagram. To obtain the result quoted in \cite{Herzog:2002pc}, where the 2-point Schwinger-Keldysh field theory propagator is given as a boundary term, with the R and L contributions subtracted from each other, we will work with the bulk action
\bea
{\cal S}&=&\bar N\int_R\sqrt{-g}\bigg(g^{\mu\nu}\partial_\mu\phi\partial_\nu\phi+m^2\phi^2+\;{\rm interactions}\bigg)\nn\\
&-&\bar N\int_L\sqrt{-g}\bigg(g^{\mu\nu}\partial_\mu\phi\partial_\nu\phi+m^2\phi^2+\;{\rm interactions}\bigg),\label{r-l}
\eea
where $\bar N$ is a supergravity normalization factor which includes the volume factor of $S^5$, which was integrated over implicitly (to be specific, $\bar N=-N_c^2/(16\pi^2 R^3)$, where we recall that $N_c$ is the number of colors in the dual ${\cal N}{=}4$ super Yang-Mills theory, and $R$ is the radius of $S^5$).

In (\ref{r-l}) the bulk right and bulk left contributions, from the respective boundaries to the event horizon, come with opposite signs {\it ab initio}. A similar action was used by Frolov and Martinez \cite{Frolov} in their construction of the action and Hamiltonian of eternal black holes in what they called ``tilted foliation''. The Hamiltonian received two opposite sign contributions from the two causally disconnected regions, namely the R and L quadrants. Frolov and Martinez \cite{Frolov} pointed out that the Fock states in the R and L regions are akin to the doubling of the degrees of freedom (physical/doubler) encountered in the real-time formalism of finite-temperature field theories. Moreover, the structure of the Hamiltonian, as $H=H_R-H_L$, is also similar to the real-time Hamiltonian, which includes a physical particle Fock space contribution and a doubler Fock space contribution.

The Kruskal coordinates $t_K$ and $x_K$ given in (\ref{kruskal}) are useful when considering global aspects of the spacetime and for defining notions such as incoming/outgoing and positive/negative frequency modes. However, these coordinates are cumbersome when it comes to solving the equations of motion for $\phi$ and computing its action. For these tasks, it is simpler to return to the coordinate $u$, in terms of which the $AdS$ boundaries are given by $u=0$. However, $u$ is only defined for one half of the extended AdS Schwarzschild spacetime, so we need to define two such coordinates, $u_L$ and $u_R$. $u_R$ is precisely our original $u$ coordinate. We have
\be
\int_R\sqrt{-g} = \int d^4x\int_0^1 du_R\sqrt{-g},\qquad \int_L\sqrt{-g} = \int d^4x \int_0^1 du_L\sqrt{-g}.
\ee
It may seem that there should be a minus sign in the second equation since $u_L$ increases with increasing $x_K$, while $u_R$ decreases with increasing $x_K$. However, $t_R$ and $t_L$ flow in opposite directions with respect to $t_K$. ($t_R$ flows roughly parallel to $t_K$, while $t_L$ is antiparallel.)

Only the boundary terms of the quadratic action ${\cal S}_0$  play a role in the computation of the 2-point function. Since the boundary is perpendicular to the $u$ direction, only the $u$-derivative terms,
\be
{\cal S}_0=\bar N\int d^4x\int_0^1 du_R \sqrt{-g}g^{uu}(\partial_u\phi)^2-\bar N\int d^4x\int_0^1 du_L\sqrt{-g}g^{uu}(\partial_u\phi)^2+...
\ee
contribute to the boundary terms. Writing
\be
\phi(x,u)=\int {d^4p\over(2\pi)^4}e^{i p\cdot x}\phi(p,u),
\ee
where $p^\mu = (E,\vec P)$, 
and integrating by parts, we obtain the boundary terms\footnote{The horizon terms have been thrown away, based on the same considerations as in \cite{Son:2002sd}, namely the zero-temperature limit is correctly reproduced by the boundary terms at $u_{L,R}=0$.}:
\bea
&&{\cal S}_0=-{\bar N\over2}\int {d^4p}\sqrt{-g}g^{uu}\phi(-p)\partial_u\phi(p)\bigg|_{u_R\to 0}+{\bar N\over2}\int {d^4p}\sqrt{-g}g^{uu}\phi(-p)\partial_u\phi(p)\bigg|_{u_L\to 0}.\label{boundaryterms}\nonumber\\
\eea
The 2-point functions are obtained by imposing the following boundary conditions on the scalar field $\phi$ \cite{Herzog:2002pc}:

i) $\phi(p,u)$ is such that in the right quadrant, at the horizon, the positive energy modes are incoming and the negative energy modes are outgoing; the left quadrant modes are then uniquely determined by analyticity;

ii)) $\phi(p,u)$ approaches two distinct functions at the two boundaries:
\be
\lim_{u_R\to 0}\phi(p,u_R)= \phi_1(p),\qquad
\lim_{u_L\to 0} \phi(p,u_L) = \phi_2(p).
\ee
With this boundary prescription, the bulk scalar field in the R and L quadrants reads
\be
\phi(p,u_a)=\phi_b(p) {\cal G}_{ba}(p,u),\label{scalarsoln}
\ee
where\footnote{The relationship between $f_k(u)$ in \cite{Herzog:2002pc} and our $F(\omega,\vec p,u)$ is complex conjugation. }
\bea
&&{\cal G}_{11}= \frac{e^{2\omega\pi}}{e^{2\omega\pi}-1}F(p,u_R)\nonumber-\frac{1}{e^{2\omega\pi }-1}F(-p,u_R),
\nonumber\\
&&{\cal G}_{21}=
 2i\frac{e^{\omega\pi}}{e^{2\omega\pi}-1}\Im F(-p,u_R) ,\nonumber\\
&&{\cal G}_{12}= -2i\frac{e^{\omega\pi}}{e^{2\omega\pi}-1}\Im F(-p,u_L),\nonumber\\
&&{\cal G}_{22}=\frac{e^{2\omega\pi}}{e^{2\omega\pi}-1}F(-p,u_L)-\frac{1}{e^{2\omega\pi}-1}F(p, u_L).
\label{bulkbndy}
\eea
The indices $a$ and $b$ take the values 1 and 2, where 1 corresponds to the R quadrant, and 2 the L quadrant. We have adopted the convention that, for example, ${\cal G}_{12}$ is a propagator\footnote{Strictly speaking, ${\cal G}_{ab}$ are scalar bulk-to-boundary 2-point Green's functions, as indicated by the use of the letter ${\cal G}$. However, we will keep referring to ${\cal G}_{ab}$ as scalar propagators, and we hope that this is not a source of confusion.} which extends from a point on the R boundary to a point in the L bulk. By identifying $F$ with the retarded bulk-to-boundary propagator, 
\be
F(p,u_R)={\cal G}_R,
\ee
then the first equation in (\ref{bulkbndy}) is consistent with our definition of the Feynman bulk-to-boundary propagator (\ref{th})
\be
{\cal G}_{11}={\cal G}_F.
\ee

Plugging these solutions into the above action and varying with respect to $\phi_1$ and $\phi_2$ yields a $2\times 2$ matrix of 2-point functions $G_{ab}$.
It is important to recall that even though the scalar field and its action were real to begin with, the boundary conditions one had to impose at the horizon break reality. The on-shell action evaluated on the solution (\ref{scalarsoln}) is therefore complex and yields complex 2-point functions
\be
G_{ab}(p_1,p_2)=-(-1)^{a+b}\frac{\delta^2 {\cal S}_0}{\delta \phi_a(p_1)\delta\phi_b(p_2)}.
\ee
More concretely, the 2-point functions are
\bea
G_{11}&=&-2\bar N\sqrt{-g} g^{uu} \frac{(e^{2\omega\pi}-1)F(p)\partial_u F(p)+ (e^{-2\omega\pi}-1)F(-p)\partial_u F(-p)}{(e^{2\omega\pi}-1)(e^{-2\omega\pi}-1)}\bigg|_{u=0},\nonumber\\
G_{12}&=&-2\bar N\sqrt{-g} g^{uu}\frac{e^{\omega \pi}}{e^{2\omega\pi}-1}\bigg(F(-p)\partial_u F(-p)-F(p)\partial_u F(p)\bigg)\bigg|_{u=0},\nonumber\\
G_{21}&=&-G_{12}^*,\nonumber\\
G_{22}&=&-G_{11}^*,
\eea
where we stripped away a momentum delta-function.
This differs slightly from the result quoted in Son and Herzog \cite{Herzog:2002pc} in that
their retarded 2-point function is $G_R=-2\bar N\sqrt{g} g^{uu}F(-p,u)\partial_u F(p,u)|_{u=0}$,
whereas we find that 
\be
G_R(p)=-2\bar N\sqrt{-g} g^{uu}F(p,u)\partial_u F(p,u)|_{u=0}.
\ee
However, since at the boundary $F(p,u)$ is normalized to 1 and the term linear in $u$ in $F(p,u)$ is insensitive to the replacement of $(\omega,\vec p)$ by $(-\omega,-\vec p)$, this difference is mostly cosmetic. Substituting (\ref{F2}), and using the small $u$ expansions in (\ref{smalluheun}),
the retarded 2-point function evaluates to\footnote{We have followed Son and Starinets by normalizing $F(u=0)=1$ and regularized $G_R(p)$ by computing it at $u=u_B\ll1$. Alternatively, we could have normalized $F(u=u_B)=1$, and found that equation (\ref{gr2pt1}) changes by the addition of $-\frac{\pi^2 N_c^2 T_H^4}{2} (\omega^2-|\vec p|^2)^2$.}
\bea
G_R(p)&=&-4\bar N(\pi T_H)^4 R^3 \left(c_1^2+c_2+2A+2c_2'+\frac{i-1}2\omega
+2\omega c_1 (1+i) + i \omega^2 +\rm{divergent}\right)\nn\\
&=&\frac{\pi^2 N_c^2 T_H^4}{4}\left(c_1^2+c_2+2A+2c_2'+\frac{(i-1)}2\omega
+2\omega c_1 (1+i) + i \omega^2 +\rm{divergent}\right),\nn\\\label{gr2pt1}
\eea
with the divergent terms given by
\bea
{\rm divergent}&=&\left(\frac{1+i}2\omega +c_1\right)\frac{1}{u_B}+2 c_2\ln(u_B)\nn\\
&=&\frac{\omega^2-|\vec p|^2}{u_B}-(\omega^2-|\vec p|^2)^2\,\ln(u_B).
\eea
The coefficient $A$ was computed in the limit of small frequency in (\ref{A small omega limit}). The divergent terms can be dealt with either by subtracting the zero-temperature result for the retarded 2-point correlator, or by employing boundary renormalization. As we will see in the next section, the finite-temperature CFT retarded 2-point can be obtained by analytically continuing the Euclidean result, yielding a natural expression in terms of a retarded and advanced supergravity mode propagator (a scalar mode in the case under consideration), just as we have obtained it from the R-L bulk action prescription, upon using energy-momentum conservation.

\subsubsection{Analytic continuation }

In the context of AdS/CFT, it has been pointed out by Gubser et al. \cite{Gubser:2008sz}   and by Iqbal and Liu \cite{Iqbal:2008by,Iqbal:2009fd} that retarded CFT correlators can be obtained from Euclidean correlators by performing an analytic continuation.
As before, only a boundary term in the classical action contributes to 2-point functions:
\be
{\cal S}_0=-\bar N\int {d^4k\over(2\pi)^4}\sqrt{g}g^{uu}F_E(-k,u)\partial_u F_E(k,u)\phi_0(-k)\phi_0(k)\Big|_{u=0}.\label{euclideanaction}
\ee
Since the integrand is singular as $u\to0$, we placed the boundary at $u=u_B\ll 1$.

Varying the Euclidean action with respect to $\phi(-k)$ and $\phi(k)$ yields the 2-point function:
\be
G_E(k)=-\bar N\sqrt{g}g^{uu}F_E(-k,u)\partial_u F_E(k,u)\Big|_{u=0}.\label{ge}
\ee
The analytic continuation from Euclidean to Minkowski space  in (\ref{ge}) will produce either the retarded or the advanced propagators in the right-hand-side of (\ref{ge}), depending on the sign of $\omega_E$. In particular, for $\omega_E>0$, this reduces to the previous expression found via the bulk action (\ref{r-l}), namely
$\bar N\sqrt{g} g^{uu}F(p,u)\partial_u F(p,u)|_{u=0}$.

\subsection{Real-time finite-temperature 3-point functions}

Finally, we come to the main result of this paper, which is the prescription for computing CFT 3-point functions at finite temperature and in real time via supergravity diagrams.

Once we have the bulk-to-boundary propagators ${\cal G}_{ab}$, it is straightforward to compute 3-point functions. These are obtained at tree level by computing the $\phi^3$ term (with coupling $\Lambda$ say) in the action (\ref{r-l}), plugging in the same solutions to the equation of motion for $\phi$ as in (\ref{scalarsoln}), and varying with respect to the boundary fields $\phi_1$ and $\phi_2$ as in (\ref{nptG}):
\be
G_{abc}=-(-1)^{a+b+c}\Lambda \bar N \bigg(\int_0^1 du_R
\sqrt{-g}{\cal G}_{a1}{\cal G}_{b1}{\cal G}_{c1}-\int_0^1 du_L \sqrt{-g} {\cal G}_{a2}{\cal G}_{b2}{\cal G}_{c2}\bigg),\label{gammaabc}
\ee
where we recall that ${\cal G}_{ab}$ were defined in (\ref{bulkbndy}).

 We would like to check if this expression for $G_{abc}$ which we have obtained from the R minus L quadrant prescription for the AdS-S supergravity action (\ref{r-l}) is correct.
A first check would be to verify whether (\ref{gammaabc})
obeys known identities for Schwinger-Keldysh 3-point functions. In particular the KMS identities for the 3-point functions read
\be
G_{abc}^*=G_{\bar a \bar b\bar c}\label{kms3},
\ee
where $\bar 1=2$ and $\bar 2=1$, and we have assumed that $\sigma=\beta/2$. Using that the bulk-to-boundary propagators ${\cal G}_{ab}$ have  the property
\be
{\cal G}_{ab}={\cal G}_{\bar a\bar b}^*,
\ee
it can be shown that, starting from (\ref{gammaabc}), the identities (\ref{kms3}) are indeed satisfied.

Yet another test which (\ref{gammaabc}) passes is to verify the largest time equation (\ref{largest3}). On the gravity side we can think of this identity as arising from summing tree-level scalar field diagrams, with three vertices on the boundary of the R quadrant and one vertex in the bulk, with each vertex being either circled or uncircled for a total of $2^4$ diagrams. As expected in a causal theory, the sum is zero.
\begin{center}
\begin{figure}[h]
\begin{center}
\includegraphics[width=13 cm]{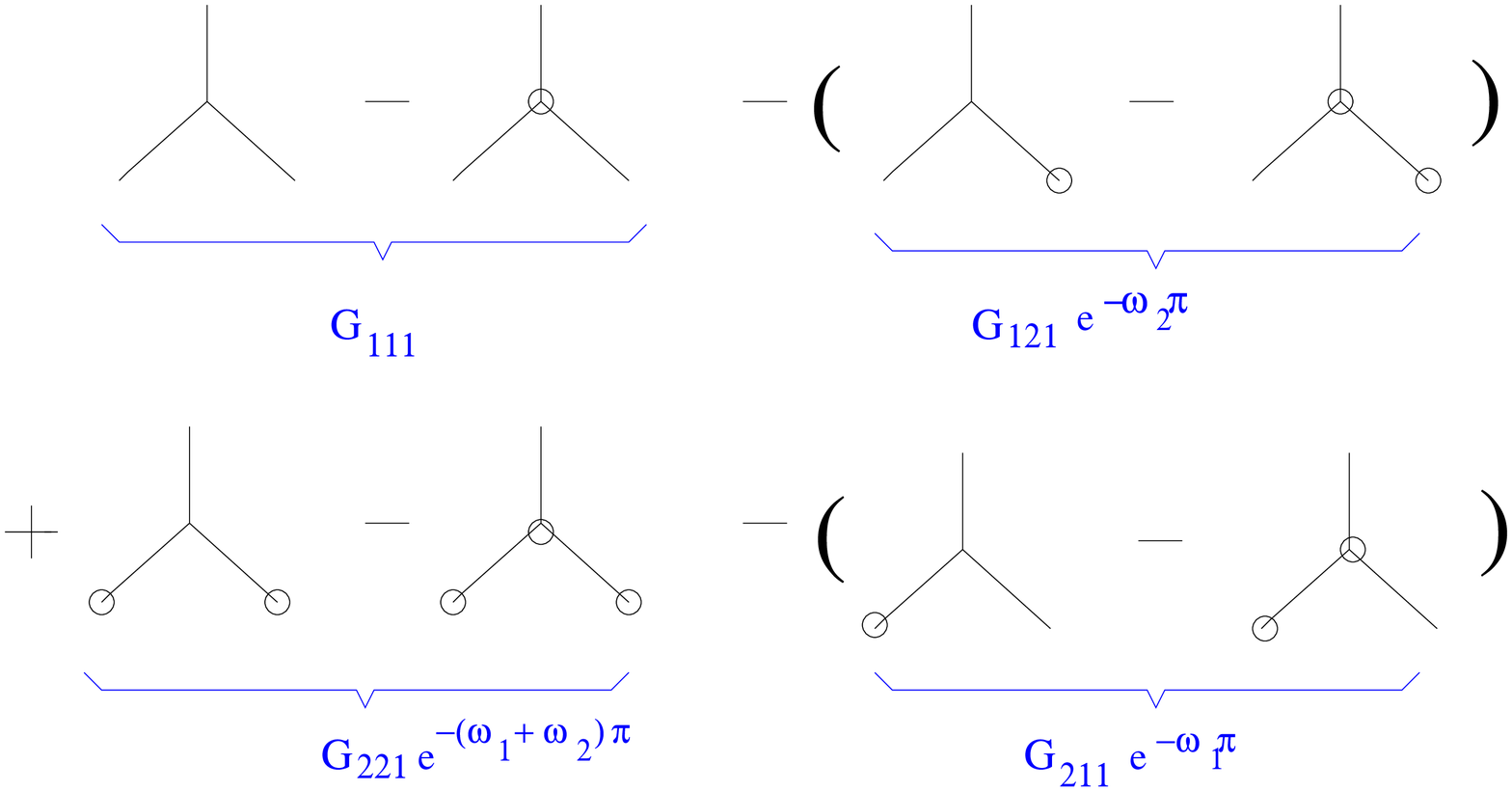}\\
Figure 6: The retarded 3-point function from circling rules
\end{center}
\end{figure}
\end{center}
Next, following \cite{kobes} we define the ``causal'' (retarded) 3-point correlator to be given by (\ref{causal3point}). As shown by Kobes \cite{kobes}, the causal n-point functions are special, because they make contact with the n-point function computed in imaginary-time formalism.  The causal real-time Green's functions are obtained from the imaginary-time Green's function by analytic continuation.

When substituting the various $G_{abc}$ in terms of bulk integrals in (\ref{causal3point}), given the multitude of terms that are added, we find that the final expression is surprisingly simple\footnote{A similar conclusion was reached in \cite{vr} regarding the retarded 3-point function.}:
\be
G_R(q,p;r)=\Lambda \bar N\delta^4(p+q+r)
\int_0^1 du \sqrt{-g} F^*(q) F^*(p) F(r)\label{3pointadsT},
\ee
where $F(r)$ is the retarded bulk-to-boundary propagator ${\cal G}_R$. In hindsight, the retarded 3-point correlator could only take the form in (\ref{3pointadsT}). That is because, by the circling rules, the retarded 3-point function can be written as a sum of R-quadrant tree-level gravity diagrams, with all vertices being either circled or uncircled with the exception of the vertex which has the largest time and which is uncircled. These are the diagrams shown in Figure 6.
Their sum reduces to a single tree-level diagram, with two advanced and one retarded bulk-to-boundary propagator. This is the only possible answer for a causal theory when computing a tree-level retarded 3-point function.

As another check on (\ref{3pointadsT}), we can take the zero-temperature limit. In this limit, the retarded/advanced  bulk-to-boundary propagators of a massless scalar field in the AdS-S background approach the AdS retarded/advanced propagators (given in Appendix C), as discussed in Section 4.2.3. The integration limits extend in the zero-temperature limit to $(0,\infty)$, and the integral in (\ref{3pointadsT}) reproduces  (\ref{t0ret3pt}). We have also verified that (\ref{3pointadsT}) can be obtained by analytically continuing the Euclidean space 3-point function.




\begin{appendix}
\section{Momentum-space 2-point CFT correlators at $T{=}0$}
We now want to find the momentum-space expression of all 2 and 3-point functions given above.
The 2-point correlator in momentum space, Euclidean signature is
\be
G_E(k)=-i\frac{\Gamma(2-\Delta-\epsilon)\mu^{-2\epsilon}}{4^{\Delta-1-\epsilon}\Gamma(\Delta)}(k^2)^{\Delta-2+\epsilon}\label{2pointE},
\ee
where $k^2=E^2+\vec k^2$, and the Fourier-transform integral was regularized by dimensional regularization. Note that if the conformal dimension $\Delta$ is a positive integer greater or equal to 2, then the 2-point will contain logs. Not surprisingly, this result coincides with the AdS/CFT computation.

In Minkowski signature, the story is a bit more complex. The Fourier transform of the scalar field 2-point Green's functions yields:
\bea
G^\pm &=& -2\pi i \theta(\pm E) \delta(E^2-\vec P^2),\\
G_{R}(p)&=& -\frac{ 1 }{-(E+i\epsilon)^2+\vec P^2},\qquad G_A(p)=G_R^*(p),\\
G_{F}(p)&=&-\frac{1}{-E^2+\vec P^2 - i\epsilon},
\eea
while the momentum-space expressions of the non-time ordered, retarded, advanced and time-ordered 2-point scalar field operator correlators are
\bea
G_{\Delta> 2}^\pm &=& -\frac{2\pi i (\Delta-1) }{(2^{\Delta-1} \Gamma(\Delta))^2}\theta(\pm E-|\vec P|) (E^2-\vec P^2)^{\Delta-2}\\
G_{R; \,\Delta> 2}(p)&=& \frac{ (\Delta-1)}{(2^{\Delta-1}\Gamma(\Delta))^2}(E^2-\vec P^2)^{\Delta-2}\ln(-(E+i\epsilon)^2+\vec P^2)\epsilon^2)\\&=&\frac{ (\Delta-1)}{(2^{\Delta-1}\Gamma(\Delta))^2}(E^2-\vec P^2)^{\Delta-2}\bigg(\ln(|-E^2+\vec P^2| \epsilon^2) - i\pi\theta(E^2-\vec P^2) sgn(E)\bigg)\nonumber\\
G_A(p)&=&G_R^*(p)\\
G_{F;\,\Delta> 2}(p)&=&\frac{(\Delta-1) }{(2^{\Delta-1}\Gamma(\Delta))^2}(E^2-\vec P^2)^{\Delta-2}\bigg(\ln(|-E^2+\vec P^2|\epsilon^2)-i\pi \theta(E^2-\vec P^2)\bigg),
\eea
where $\Lambda=1/\epsilon$ was used as a UV cut-off to regularize some of the integrals, and $\Delta$ was assumed to be an integer, for concreteness.

Note that the analytic continuation of the Euclidean signature 2-point function, with $E\to -i(E\pm i\epsilon) $ yields the retarded/advanced 2-point correlators.

\section{Momentum-space real-time 3-point correlators at $T{=}0$}
Here we present the intermediate steps leading to the results quoted in Section 2.2. We begin with the Euclidean 3-point correlator in position space, and Fourier transform. Since the integral is not convergent in $d=4$, we use dimensional regularization, with $d=4-2\epsilon$:
\bea
&&\!\!\!\!\!\! \Gamma(\delta)^3  i^2G_E(k_1,k_2,k_3)=\Gamma(\delta)^3 \bigg(
\prod_{i=1}^{3}\int d^{d} x_i \, {e^{-i k_i\cdot x_i}}\bigg)\frac{1}{x_{12}^{2\delta} x_{23}^{2\delta} x_{31}^{2\delta}}\nonumber\\
&=&\bigg(\prod_{i=1}^{3}\int d^{d} x_i  \int_0^\infty d s_i \, s_i^{\delta-1} e^{-i k_i \cdot x_i} \bigg)e^{-s_1 x_{12}^2} e^{-s_2 x_{23}^2} e^{-s_3 x_{31}^2}\nonumber\\
&=&\bigg(\prod_{i=1}^3 \int_0^\infty d s_i \, s_i^{\delta-1} \bigg)e^{-\frac{k_1^2 s_2+k_2^2 s_3+k_3^2 s_1}{4(s_1 s_2 + s_2 s_3 + s_3 s_1)}} (2\pi)^{d} \pi^{{d}} \frac{\delta^{d}(k_1+k_2+k_3)}{( s_1 s_2 + s_2 s_3 + s_3 s_1)^{\frac{d}{2}}}\nonumber\\
&=&(2\pi^2)^{d} \delta^{4-2\epsilon}(k_1+k_2+k_3)\bigg(\prod_{i=1}^3 \int_0^\infty d u_i \,
{u_i}^{\delta-1} e^{-\frac{k_i^2 u_i}{4}} \bigg)
\bigg(\frac{1}{u_1 u_2 u_3(\frac{1}{u_1}+\frac{1}{ u_2} + \frac{1}{ u_3})}\bigg)^{3\delta - \frac{d}{2}}\nonumber\\
&=&(2\pi^2)^{d}\frac{\delta^{d}(k_1+k_2+k_3)}{\Gamma(3\delta-\frac{d}{2})}\int_0^\infty ds \bigg(\prod_{i=1}^3 \int_0^\infty d u_i \, u_i^{-2\delta-1+\frac{d}{2}}e^{-\frac{k_i^2 u_i}{4}}e^{-\frac{s}{u_i}}\bigg) s^{3\delta-\frac{d}{2}-1}
\nonumber\\
&=&(2\pi^2)^{d}\frac{\delta^{d}(k_1+k_2+k_3)}{\Gamma(3\delta-\frac{d}{2})}\int_0^\infty dz
\bigg(\frac{z}{2}\bigg)^{6\delta-1-d}
\bigg(\prod_{i=1}^3 \int_0^\infty d u_i \,e^{-\frac{k_i^2 u_i}{4}}
e^{-\frac{z^2}{4 u_i}} u_i^{-2\delta-1+\frac{d}{2}}\bigg)\nonumber\\
&=&2^{10-3\Delta{-3\epsilon}}(2\pi^2)^{d} \frac{\delta^{d}(k_1+k_2+k_3)}{\Gamma(\frac{3\Delta}2-\frac{d}{2})}\int_0^\infty \frac{dz}{z^{d+1}}
\prod_{i=1}^3 (\sqrt{k_i^{2}})^{\Delta-\frac{d}{2}}z^{\frac{d}{2}}
K_{\Delta-\frac{d}{2}}(\sqrt{k_i^2}z)
\nonumber\\
\eea
which is the expression we gave in (\ref{3pointE}).

Next, we compute the Fourier transform of (\ref{circ3}) first in the kinematics 1,2 incoming and 3 outgoing, that is $E_1<0, E_2<0, E_3>0$.
The steps taken next follow closely  the path used for the Euclidean 3-point correlator. We could have chosen to regularize the integrals using dimensional regularization, as we did before. However, here we chose a different regularization to show how the  momentum-space correlator can be obtained via a 5-dimensional AdS integral. At each step in our manipulations we will make sure that we appropriately insert convergence factors, such that the integrals are well defined.

To begin, we use again Schwinger parameters to write the denominators of the first term in (\ref{circ3}), which is the only term that contributes in the chosen kinematics. For example,
\be
\bigg(\frac{1}{-t^2+\vec x^2+i\epsilon}\bigg)^\delta=\frac{(-i)^\delta}{\Gamma(\delta)}\int_0^\infty ds \,e^{-s\epsilon+is(\vec x^2-t^2)} s^{\delta-1}.
\ee
The integrals over the position-space coordinates are of the type
\be
\int_{-\infty}^\infty dz e^{i\alpha z^2}=\sqrt{\frac{\pi}{|\alpha|}}e^{i\mathrm{sgn}(\alpha)\pi/4}, \qquad \alpha \in {\bf R}.
\ee
The intermediate result, where we still have to perform the integral over the three Schwinger parameters, is very similar to what we have encountered in Euclidean signature:
\bea
i^2 G_R(p_1,p_2;p_3)&=&\frac{(2\pi^2)^4(-i)^{3\delta}}{\Gamma(\delta)^3}\delta^4(p_1+p_2+p_3) \prod_{i-1}^3 \bigg( \int_0^\infty \frac{du_i}{u_i^{2\delta-1}}
e^{-\frac{\tilde\epsilon u_i}4-\frac{\hat\epsilon}{u_i}}\bigg)\times
\nonumber\\
&&\frac{1}{(\frac{1}{u_1}+\frac{1}{u_2}+\frac{1}{u_3})^{3\delta-2}}
\exp\left(-\frac i4(u_1 \,p_1^2+u_2 \,p_2^2+u_3\, p_3^2)\right),\label{regconv}
\eea
where $p_1^2=-E_1^2+\vec P_1^2$ etc. The integrand in (\ref{regconv}) is properly regularized for both $u_i=0$ and for large values of $u_i$.
To deal with the $1/(\frac 1{u_1}+
\frac{1}{u_2}+\frac{1}{u_3})$ factor, we introduce another Schwinger-type parameter, but in a slightly different fashion than previously
\be
\frac{1}{(a+i\epsilon')^n}=\frac{2 e^{-\frac{in\pi}{2}}}{\Gamma(n)}\int_0^\infty dz
z^{2n-1}e^{i(a+i\hat \epsilon')z^2},\qquad a\in {\bf R}.
\ee
 The three Schwinger parameter $u_i$ integrals are of the type
\be
\int_0^\infty du_i e^{\frac{i u_i(-p_i^2+ i\tilde\epsilon)}{4}+\frac{i(z^2+i\hat \epsilon)}{u_i}}u_i^{-\Delta+1}=2^{3-\Delta}\left(
\frac{i p_i^2+\tilde
\epsilon}{-i z^2+ \hat \epsilon}\right)^{\frac{\Delta-2}{2}}
K_{\Delta-2}(\sqrt{(i p_i^2+\tilde \epsilon)(-i z^2+ \hat \epsilon)}).\label{usefulK}
\ee
In what follows we set the convergence factor $\hat \epsilon=0$.
Thus we find that the retarded 3-point function in momentum space can be written as
\bea
G_R(p_1,p_2;p_3)&=&\frac{(2\pi)^{8}}{2^{3\Delta-6} \Gamma(\Delta/2)^3\Gamma(3\Delta/2-2)}\delta^4(p_1+p_2+p_3)\int_0^\infty \frac{dz}{z^5}
\times \nonumber\\
&&\bigg[ z^2 (-E_1^2+\vec p_1^2-i\epsilon)^{\frac\Delta 2-1} K_{\Delta-2}(z\sqrt{-E_1^2+\vec p_1^2-i\epsilon})
\nonumber\\
&&  z^2 ( -E_2^2+\vec p_2^2-i\epsilon)^{\frac\Delta 2-1} K_{\Delta-2}(z\sqrt{-E_2^2+\vec p_2^2-i\epsilon})\nonumber\\
&&  z^2 ( -E_3^2+\vec p_3^2+i\epsilon)^{\frac\Delta 2-1} K_{\Delta-2}(z\sqrt{-E_3^2+\vec p_3^2+i\epsilon})
\bigg].
\label{t0ret3pt00}
\eea
A convergence factor $e^{-\epsilon' z^2}$ was also dropped since the integrand is convergent at large $z$. To simplify notation we removed the tilde from  $\tilde\epsilon$.

From the largest time equation we learn that the causal n-point function is real:
\be
G_R(\{x\})=G_R^*(\{x\})
\ee
which implies for the momentum-space causal retarded n-point function
\be
G_R^*(\{p\})=G_R(\{-p\}).
\ee
In particular, this gives us the causal retarded 3-point function for the energy configuration $E_{1,2}>0$ and $E_3<0$, and leads us to the following expression for $G_R(p_1,p_2;p_3)$ for general kinematics
\bea
G_R(p_1,p_2;p_3)&=&\frac{(2\pi)^{8}}{2^{3\Delta-6} \Gamma(\Delta/2)^3\Gamma(3\Delta/2-2)}\delta^4(p_1+p_2+p_3)\int_0^\infty \frac{dz}{z^5}
\times \nonumber\\
&&\bigg[ z^2 (-(E_1-i\epsilon)^2+\vec p_1^2)^{\frac\Delta 2-1} K_{\Delta-2}(z\sqrt{-(E_1-i\epsilon)^2+\vec p_1^2})
\nonumber\\
&& z^2 ( -(E_2-i\epsilon)^2+\vec p_2^2)^{\frac\Delta 2-1} K_{\Delta-2}(z\sqrt{-(E_2-i\epsilon)^2+\vec p_2^2})\nonumber\\
&& z^2 ( -(E_3+i\epsilon))^2+\vec p_3^2)^{\frac\Delta 2-1} K_{\Delta-2}(z\sqrt{-(E_3+i\epsilon)^2+\vec p_3^2})
\bigg].\nn\\
\label{t0ret3pt0}
\eea

\section{The retarded bulk-to-boundary scalar propagator in AdS space in Poincar\'{e} coordinates}

In this section we briefly discuss the causal properties of the retarded/advanced bulk-to-boundary scalar propagator in 5-dimensional AdS
\be
{\cal G}_R={\cal C}\, z^2 ( -(E+i\epsilon))^2+\vec p^2)^{\frac\Delta 2-1} K_{\Delta-2}(z\sqrt{-(E+i\epsilon)^2+\vec p^2}), 
\ee
where ${\cal C}$ is a real-valued normalization constant.
The simpler route to constructing the retarded propagator in position space is to first compute the Feynman propagator
\be
\!\!
{\cal G}_F(E,\vec p,u)={\cal G}_R(E,\vec p,u) \;{\rm if}\; E>0,\qquad {\rm and}\qquad {\cal G}_F(E,\vec p,u)={\cal G}_R^*(E,\vec p,u)\;{\rm if}\; E<0
\ee
which leads to
\be
{\cal G}_F={\cal C}\, z^2 ( -E^2 +\vec p^2 - i\epsilon )^{\frac\Delta 2-1} K_{\Delta-2}(z\sqrt{-E^2+\vec p^2 - i\epsilon}).
\ee
Then the Fourier transform of the Feynman propagator can be computed with the help of (\ref{usefulK})
\be
{\cal G}_F={\cal C}\frac{i2^{\Delta-3}\Gamma(\Delta)}{\pi^2}\frac{z^\Delta}{(-t^2+\vec x^2+z^2+i\epsilon)^\Delta}.
\ee
This is the expression we might have gotten starting from the Feynman bulk-to-boundary scalar propagator computed in Euclidean AdS, after performing the usual Wick rotation and analytic continuation.

Hence the retarded propagator in position space obtained from
\be
{\cal G}_R=\theta(t)({\cal G}_F+{\cal G}_F^*)
\ee
is causal, and has support only on the forward light cone.

\section{The retarded 3-point momentum-space correlator with $\Delta{=}2$ and at $T{=}0$}

In this appendix we give a closed-form analytic answer for the retarded 3-point CFT correlator, of three scalar operators with conformal dimension $\Delta=2$.
Again we begin from the position-space (\ref{retarded3point}), and we perform the Fourier transform as follows:
\bea
\!\!\!\!\!\!&&\!\!\!\!\!\!i^2G_R(p_1,p_2;p_3) =
\prod_{i=1}^3 \int {d^4 x_i}  e^{-i p_i x_i} i^2 G_R (x_1, x_2, x_3)\nonumber\\
&=&\int {d^4 x_2} e^{-i \sum_{i=1}^3 p_i x_2} \int
{d^4 x_{12}} e^{-ip_1 x_{12}} \int{ d^4 x_{23}} e^{ip_3 x_{23}} \int {d^4 x_{31}} \int \frac{d^4 k}{(2\pi)^4} e^{-ik(x_{12}+x_{23}+x_{31})} G_R (x_{1}, x_{2}; x_{3}) \nonumber\\
&=& \delta^4 (p_1+p_2+p_3)
\int {d^4 k}\int_{-\infty}^\infty
\frac{dz_1}{2\pi i(z_1-i\epsilon)} \int_{-\infty}^\infty
\frac{dz_2}{2\pi i(z_2-i\epsilon)} \int{ d^4 x_{12}}  \int{ d^4 x_{23}}  \int{ d^4 x_{31}} \nonumber\\
&&\bigg[e^{i (E_1 + z_1+ \Omega )t_{12}} e^{-i(\vec P_1+\vec K) \vec x_{12}}
e^{-i(E_3 - \Omega) t_{23}} e^{i(\vec P_3 - \vec K)\vec x_{23}}
e^{i(\Omega + z_2)t_{31}} e^{-i \vec K \vec x_{31}}\nonumber\\&&
\bigg(\frac{1}{x_{12}^2 + i\epsilon t_{12} }\frac{1}{x_{23}^2-i\epsilon t_{23}} -c.c\bigg)\bigg(  \frac{1}{x_{31}^2+i\epsilon t_{31}} - c.c.\bigg)
\nonumber\\
&&+e^{i (E_1 - z_1+ \Omega )t_{12}} e^{-i(\vec P_1+\vec K) \vec x_{12}}
e^{-i(E_3 +z_2- \Omega) t_{23}} e^{i(\vec P_3 - \vec K)\vec x_{23}}
e^{i \Omega t_{31}} e^{-i \vec K \vec x_{31}}\nonumber\\
&&\bigg(\frac{1}{x_{12}^2 - i\epsilon t_{12} }\frac{1}{x_{31}^2+i\epsilon t_{31}} -c.c\bigg)\bigg(  \frac{1}{x_{23}^2-i\epsilon t_{23}} - c.c.\bigg)
\bigg].\nonumber\\
\eea
Next substitute the Fourier transform of the functions depending on $x_{12}, x_{23}, x_{31}$, and use that these are of the type
\be
\theta(\pm E) \delta (E^2-\vec P^2) = \frac{\delta(|\vec P|\mp E)}{2|\vec P|}
\ee
to perform the integrals over $z_1, z_2$ and $\Omega$. The remaining $d^3K$ integral can be simplified by going to a Lorentz frame where $\vec P_3$ is vanishing:
\bea
&&\!\!\!\!\!\!i^2G_R(p_1,p_2;p_3) =\frac{(4\pi^3)^3}{(2\pi i)^2}\int d^3 \vec K \bigg(\frac{1}{|\vec P_1 + \vec K|+|\vec K|+E_1+E_3 + i\epsilon}\,\frac{-2|\vec K|}{(|\vec K|+E_3+i\epsilon)^2 - \vec |K|^2}\nonumber\\
&&-\frac{1}{-|\vec P_1 + \vec K|-|\vec K|+E_1+E_3 + i\epsilon}\,\frac{-2|\vec K|}{(-|\vec K|+E_3+i\epsilon)^2 - |\vec K|^2}\nonumber\\
&&+\frac{1}{|\vec P_1+\vec K|+|\vec K|-E_1+i\epsilon}\,\frac{-2 |\vec K|}{(|\vec K|+E_3+i\epsilon)^2-|\vec K|^2}\nonumber\\
&&-\frac{1}{-|\vec P_1+\vec K|-|\vec K|-E_1+i\epsilon}\,\frac{-2 |\vec K|}{(-|\vec K|+E_3+i\epsilon)^2-|\vec K|^2}\bigg)\frac{1}{|\vec K|^2|\vec P_1+\vec K|}
\frac{1}{E_3+i\epsilon}.
\eea
The 3-dimensional integral over $\vec K$ is performed by using spherical coordinates, with $\vec P_1$ aligned with the z-axis. The integral over $\phi$ is trivial and yields a factor of $2\pi$. The remaining integrals over $|\vec K|$ and $\theta$ are re-expressed as integrals over $|\vec K|$ and $|\vec P_1 + \vec K|$. The Jacobian of this change of variable is
\be
\bigg| \frac{\partial (|\vec K|,  \cos\theta)}{\partial (|\vec K|,  |\vec P_1+ \vec K|)}\bigg|=\frac{|\vec P_1 +\vec K|}{|\vec P_1| | \vec K|},
\ee
where $|\vec K+\vec P_1|$ is integrated from $||\vec K|-|\vec P_1||$ to $|\vec K|+|\vec P_1|$. Lastly the integral over $|\vec K +\vec P_1|$ is easily evaluated to a log, and the final expression of the retarded 3-point function is
\bea
i^2G_R(p_1,p_2;p_3) &=& \frac{(4\pi^3)^3 }{(2\pi i)^2}
\frac{4\pi}{|\vec P_1| }
\int_0^\infty d|\vec K|\times\nn\\
&&\bigg[-\frac{1}{E_3+i\epsilon}
\bigg(\frac{1}{E_3+2|\vec K|+i\epsilon}
\ln\frac{2 |\vec K| + |\vec P_1| + E_1+E_3+i\epsilon}
{||\vec K|-|\vec P_1||+|\vec K|+E_1+E_3+i\epsilon}
\nonumber\\
&&+\frac{1}{E_3 - 2|\vec K| +i\epsilon}
\ln\frac{-2|\vec K|-|\vec P_1| + E_1+E_3+i\epsilon}
{-||\vec K|-|\vec P_1||-|\vec K|+E_1+E_3+i\epsilon}\bigg)\nonumber\\
&&-\frac{1}{E_3+i\epsilon}
\bigg(\frac{1}{E_3+2|\vec K|+i\epsilon}
\ln\frac{2|\vec K|+|\vec P_1|-E_1+i\epsilon}
{||\vec K|-|\vec P_1||+|\vec K|-E_1+i\epsilon}\nonumber\\
&&+\frac{1}{E_3-2|\vec K|+i\epsilon}
\ln\frac{-2|\vec K|-|\vec P_1|-E_1+i\epsilon}
{-||\vec K|-|\vec P_1||-|\vec K|-E_1+i\epsilon}\bigg)\bigg].
\eea
With some effort, the integral over $|\vec K|$ can be performed analytically:
\bea
i^2G_R(p_1,p_2;p_3)&=&\frac{(4\pi)^3 (2\pi)}{(2\pi i)^2|\vec P_1|(E_3+i\epsilon)}\bigg[2Li_2\bigg(\frac{-E_2+|\vec P_1|+i\epsilon}{E_1+|\vec P_1|-i\epsilon}\bigg)
-2Li_2\bigg(\frac{E_2+|\vec P_1|-i\epsilon}{-E_1+|\vec P_1|+i\epsilon}\bigg)\nonumber\\
&+&\frac 12\ln^2\bigg(-\frac{-E_2+|\vec P_1|+i\epsilon}{E_1+|\vec P_1|-i\epsilon}\bigg)
-\frac 12\ln^2\bigg(-\frac{E_2+|\vec P_1|-i\epsilon}{-E_1+|\vec P_1|+i\epsilon}\bigg)\nonumber\\
&+&
\frac 12\ln^2\bigg(\frac{E_2+|\vec P_1|-i\epsilon}{E_1+|\vec P_1|-i\epsilon}\bigg)
-\frac 12\ln^2\bigg(\frac{-E_2+|\vec P_1|+i\epsilon}{-E_1+|\vec P_1|+i\epsilon}\bigg)\bigg],
\eea
where we recall that we went to a special Lorentz frame such that $\vec P_3=0$. We have checked that in this frame, our result coincides with eqn. (28) of \cite{Bailey:2008ib}. The final form of the Fourier-transformed retarded 3-point function, with $p_3$ being the momentum flowing in the vertex with the largest time, is obtained by covariantizing the previous expression with the help of the following relations
\be
|\vec P_1|=\sqrt{\frac{p_1^2 p_2^2 - (p_1 p_2)^2}{p_3^2}}, \qquad E_1^2=-\frac{(p_1 p_3)^2}{p_3^2}, \qquad E_2^2 = -\frac{(p_2 p_3)^2}{p_3^2}.
\ee

\section{ The Heun's Function}

Heun functions arise as a generalization of the hypergeometric function, and are defined as solutions of the Fuchsian differential equation \cite{book1,book2,192}:
\be
\frac{d^2 H}{dz^2} + \bigg(\frac{\gamma}{z}+\frac{\delta}{z-1}+\frac{\epsilon}{z-d}\bigg)\frac{dH}{dz}+\frac{\alpha\beta z-q}{z(z-1)(z-d)}H=0, \qquad \alpha\beta,\gamma,\delta,\epsilon\in C.
\ee
This differential equation has 4 regular singular points: $0,1,d,\infty$. Regularity at infinity is insured provided that
\be
\epsilon=\alpha+\beta-\gamma-\delta+1.
\ee
The characteristic exponents at the singular points are: $0, 1-\gamma$ for $z=0$; $0,1-\delta$ for $z=1$; $0,1-\epsilon$ for $z=d$, and $\alpha,\beta$ for $z=\infty$.

The Heun's function is further defined to be normalized to 1 at z=0:
\be
Hl(d,q,\alpha,\beta,\gamma,\delta;z)\bigg|_{z=0}=1;\qquad \frac{d}{dz} Hl(d,q,\alpha,\beta,\gamma,\delta;z)\bigg|_{z=0}=\frac{q}{\gamma d}.
\ee
In the vicinity of $z=0$, the Heun's function is analytic and can be expressed in terms of a locally convergent series expansion $\sum_{k\geq 0}c_{k}z^k$, where the coefficients $c_k$ are obtained recursively
\bea
&&(k+1)(k+\gamma) d c_{k+1}-\bigg[k\bigg((k+\gamma+\delta-1)d+(k+\gamma+\epsilon-1)\bigg)+q \bigg]c_k\nonumber\\
&&+(k+\alpha-1)(k+\beta-1)c_{k-1}=0\nonumber\\
&&c_{-1}=0, c_0=1. \label{c2}
\eea
The series expansion breaks down when $k$ is a negative integer, and the Heun's function will exhibit logarithmic behaviour. This case will be further investigated, as it is relevant for the physical problem we are considering.

Based on the characteristic exponents at $z=0$, there are two independent solutions of the Heun differential equation. One is $Hl(d, q,\alpha,\beta,\gamma,\delta;z)$, and the other is $z^{1-\gamma} Hl(d,q-(\gamma-1)(\delta d+\epsilon),\beta-\gamma+1,\alpha-\gamma+1,2-\gamma,\delta;z)$. Notice that for the latter function, having $\gamma$ be a negative integer poses no special problems. Then, we can search for a particular combination of these two independent solutions such that the poles in $\gamma$ cancel. In particular, for $\gamma=-1$, the linear combination
\be
\lim_{\gamma\to-1}\bigg(Hl(d,q,\alpha,\beta,\gamma,\delta, z)-\frac{c_2}{\gamma+1}\,z^{1-\gamma}Hl(d,q-
(\gamma-1)(\delta d+\epsilon),\beta-\gamma+1,\alpha-\gamma+1,2-\gamma,\delta;z)\bigg)\label{improved},
\ee
where $c_2$ is obtained from (\ref{c2})
\be
c_2=\frac{1}{2d}\bigg[\bigg((\gamma+\delta)d+\gamma+\epsilon+q\bigg)\frac{q}{\gamma d}-\alpha\beta\bigg]\bigg|_{\gamma=-1},
\ee
is well defined near $z=0$. In (\ref{improved}), the limit is taken with $d,\alpha,\beta,\delta,z$ held fixed and with $\epsilon$ treated as a function of $\gamma$ accordining to its defining expression $\epsilon(\gamma)=\alpha+\beta-\gamma-\delta+1$. The leading term from the second Heun function will cancel the $z^2/(\gamma+1)$-terms from the series expansion of the first Heun's function. The logarithmic behaviour of (\ref{improved}) arises from the prefactor $z^{1-\gamma}/(\gamma+1)$, written as $\exp[(2-(\gamma+1))\ln(z)]/(\gamma+1)\sim z^2 (1/(\gamma+1) - \ln(z) + {\cal O}(\gamma+1))$.

The series $\sum_{k\geq 0}c_k z^k$ converges for $|z|<\min(1,|d|)$. However, through analytic continuation it is possible to extend the Heun's function on the complex plane. There are 192 known local solutions of the Heun differential equation \cite{192} which allow the Heun function to be extended in different regions of parameter space. For example,
\be
Hl(d,q,\alpha,\beta,\gamma,\delta;z)=(1-z)^{1-\delta}Hl(d,q-(\delta-1)\gamma d,\beta-\delta+1,\alpha-\delta+1,\gamma,2-\delta;z).
\ee
Or, one can instead choose to express the solution of the Heun differential equation as an analytic expansion around $z=1$ instead of $z=0$:
\be
Hl(1-d,-q+\alpha\beta,\alpha,\beta,\delta,\gamma;1-z)
\ee
is the solution with characteristic exponent 0, and normalized to 1, while
\be
(z-1)^{1-\delta}Hl(1-d,-q+(\delta-1)\gamma d+(\beta-\delta+1)(\alpha-\delta+1),\beta-\delta+1,\alpha-\delta+1,2-\delta,\gamma;1-z)\ee
is the solution with charactersitic exponent $1-\delta$. The complete list of the 192 different {\it local} solutions of the Heun function can be found in \cite{192}.

\end{appendix}


\end{document}